\pdfoutput=1
\documentclass[preprint,10pt]{aastex}
\usepackage{ifpdf}
\usepackage[utf8]{inputenc}
\usepackage{tablefootnote}
\usepackage{hyperref}
\usepackage{natbib}

\usepackage{amsmath}

\title{New Pleiades Eclipsing Binaries and a Hyades Transiting System Identified by K2}

\author{Trevor J.\ David\altaffilmark{1,2}, Kyle E.\ Conroy\altaffilmark{3}, Lynne A.\ Hillenbrand\altaffilmark{1}, Keivan G.\ Stassun\altaffilmark{3,4}, John Stauffer\altaffilmark{5}, Luisa M.\ Rebull\altaffilmark{5}, Ann Marie Cody\altaffilmark{6},
Howard Isaacson\altaffilmark{7}, 
Andrew W.\ Howard\altaffilmark{8}, Suzanne Aigrain\altaffilmark{9}
}

\altaffiltext{1}{Department of Astronomy, California Institute of Technology, Pasadena, CA 91125, USA}
\altaffiltext{2}{NSF Graduate Research Fellow}
\altaffiltext{3}{Department of Physics \& Astronomy, Vanderbilt University, Nashville, TN 37235, USA; {\tt kyle.conroy@vanderbilt.edu}}
\altaffiltext{4}{Department of Physics, Fisk University, Nashville, TN 37208, USA}
\altaffiltext{5}{Spitzer Science Center, California Institute of Technology, Pasadena, CA 91125, USA}
\altaffiltext{6}{NASA Ames Research Center, Mountain View, CA 94035, USA}
\altaffiltext{7}{Department of Astronomy, University of California, Berkeley CA 94720, USA}
\altaffiltext{8}{Institute for Astronomy, University of Hawai`i at M\={a}noa, Hilo, HI 96720-2700, USA}
\altaffiltext{9}{Department of Physics, University of Oxford, Keble Road, Oxford OX1 3RH, UK}

\newcommand{\kms}{km~s$^{-1}$}
\newcommand{\teff}{$T_\text{eff}$}
\newcommand{\msun}{M$_\odot$}
\newcommand{\rsun}{R$_\odot$}
\newcommand{\mjup}{M$_\mathrm{Jup}$}
\newcommand{\rjup}{R$_\mathrm{Jup}$}

\newcommand{\jktebop}{{\sc jktebop}}

\begin{document}

\begin{abstract}
We present the discovery in Kepler's $K2$ mission observations and our follow-up radial velocity observations from Keck/HIRES for four eclipsing binary (EB) star systems in the young benchmark Pleiades and Hyades clusters.  Based on our modeling results, we announce two new low mass ($M_{tot} < 0.6$ \msun) EBs among Pleiades members (HCG 76 and MHO 9) and we report  on two previously known Pleiades binaries that are also found to be EB systems (HII 2407 and HD 23642). We measured the masses of the binary HCG 76 to $\lesssim$2.5\% precision, and the radii to $\lesssim$4.5\% precision, which together with the precise effective temperatures yield an independent Pleiades distance of 132$\pm$5 pc.  We discuss another EB towards the Pleiades that is a possible but unlikely Pleiades cluster member (AK II 465). The two new confirmed Pleiades systems extend the mass range of Pleiades EB components to 0.2--2 \msun. Our initial measurements of the fundamental stellar parameters for the Pleiades EBs are discussed in the context of the current stellar models and the nominal cluster isochrone, finding good agreement with the stellar models of Baraffe et al (2015) at the nominal Pleiades age of 120 Myr. 

Finally, in the Hyades, we report a new low mass eclipsing system (vA 50) that was concurrently discovered and studied by \cite{mann2016}. We confirm that the eclipse is likely caused by a Neptune-sized transiting planet, and with the additional radial velocity constraints presented here we improve the constraint on the maximum mass of the planet to be $\lesssim$1.2~\mjup.
\end{abstract}

\section{Introduction}

Clusters provide a unique opportunity to study stellar evolution by assuming coevality among their members.  Eclipsing Binaries (EBs) have historically been used as a primary tool to measure masses, radii, and temperatures of stars.  Combining these together, EBs in clusters are essential to calibrating these relations. Furthermore, EBs in clusters can be used to directly determine the distance to the cluster, providing a distance determination independent of parallax \citep[see e.g.][]{milone2013}.

The \emph{Kepler} space satellite provided unprecedented precision photometry for $\sim$150,000 stars over a 4~year mission.  With a primary purpose to discover earth-like planets, \emph{Kepler} also allowed identification of nearly 2500 EBs \citep{prsa2011, slawson2011, kirk2016}.
Now that \emph{Kepler's} primary mission has reached an end, its repurposed K2 mission is now observing fields along the ecliptic with similar precision in $\sim$80~d timespans called ``campaigns'' \citep{howell2014}.   K2 has already resulted in the discovery of over 100 EBs \citep{conroy2014, lacourse2015, armstrong2015}. 

The K2 Campaign 4 included the Pleiades and the Hyades, two of the most well-studied clusters in the literature, providing a unique opportunity to identify and characterize future benchmark EBs at moderately young ages. The $K2$ Campaign 4 pointing encompassed more than 900 confirmed or candidate members of the Pleiades and 80 confirmed or candidate members of the Hyades. The field was monitored continuously between UT 2015-02-08 and UT 2015-04-20\footnote{Data release notes are available at \url{http://keplerscience.arc.nasa.gov/K2/C4drn.shtml}}.

The canonical age of the Pleiades cluster is $\tau$ = 125 $\pm$ 8 Myr, measured using the lithium depletion boundary technique \citep{stauffer1998}. A more recent analysis by \cite{dahm2015} using the same method and updated evolutionary models favors a slightly younger age of $\tau$ = 112 $\pm$ 5 Myr, but is statistically consistent with the canonical value above. The distance to the Pleiades was the subject of a long-term controversy due to discrepant parallaxes measured by the $Hipparcos$ satellite, but several independent studies have since resolved this issue; the best current estimate of $d$ = 136.2 $\pm$ 1.2 pc comes from very long baseline radio interferometry \citep{melis2014}. The Pleiades age is such that the lowest mass members (i.e., later than a spectral type of K2, roughly) are still contracting down to the main sequence, while the intermediate mass stars are steadily burning hydrogen, and the highest mass members have begun evolution off of the main sequence. Thus, the Pleiades represents a critical test for any stellar evolution model that aims to reproduce the fundamental parameters of stars from the pre-main sequence to post-main sequence phases of evolution. Fundamental calibrators, such as benchmark EBs, across a large range in mass are needed to place stringent constraints on these models.

The traditional Hyades age and distance are $\tau$ = 625 $\pm$ 50 Myr and $d$ = 46.34 $\pm$ 0.27 pc \citep{perryman1998}, though more recent analyses suggest a substantially older age $\sim$800 Myr \citep{david2015,brandt2015}. Unlike the Pleiades, all Hyades age estimates result from Hertzsprung-Russell diagram (HRD) analysis. EBs may help to resolve the age disagreement for this cluster, which serves as a critically important benchmark for many stellar evolution studies. 

\begin{figure}[ht!]
\centering
\includegraphics[width=0.4\textwidth]{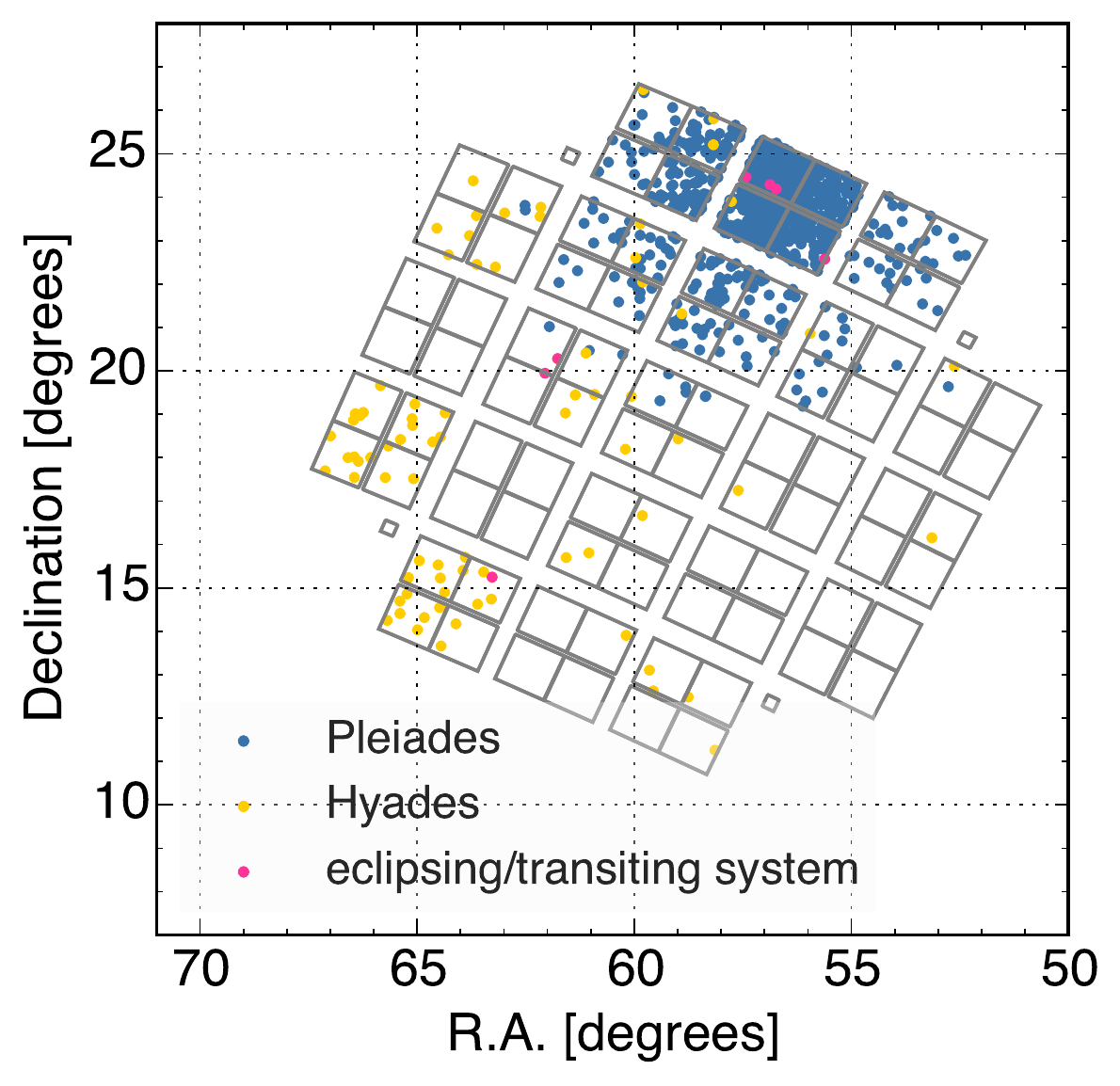}
\caption{$K2$ Campaign 4 pointing (grey)
with observed Pleiades and Hyades members overlaid. Eclipsing or transiting systems discussed in this paper are indicated by pink points.
}
\label{fig:k2fov}
\end{figure}

With its high-precision and high-cadence photometry for targets covering large portions of the Pleiades and Hyades clusters (Fig.~\ref{fig:k2fov}), K2 serves as a perfect opportunity to identify and characterize EBs in these clusters and both test and refine isochrone models, particularly the pre-main sequence (PMS) locus at 125 Myr for the Pleiades and the zero-age main sequence (ZAMS) locus for the Hyades \citep{schiller1987}. 

Before K2 there were two EBs known in the Hyades.  HD 27130 is a $\sim5.6$ day, $\sim1.8 M_\odot$ system \citep{mcclure1980,mcclure1982,schiller1987} but was not observed by K2.  
V471 Tauri \citep{guinan2001, vaccaro2015} is a $\sim0.5$ day main sequence - white dwarf binary
and was a K2 target (EPIC 210619926) but is not re-analyzed in this work. 

In the Pleiades cluster, before K2 there was only a single known EB: HD 23642 \citep{torres2003}.  The EB aspect of an additional, previously known, binary HII 2407 was recently discovered from K2 and presented in detail by \citet{david2015a}. The Pleiades EB population is valuable to establish, given the cluster age, and particularly so at low masses given the rarity of fundamental calibrators at the lowest stellar masses at any age \citep{stassun2014}. 

Here we present two new low mass EBs with certain membership in the Pleiades, one solar-mass EB with possible membership in the Pleiades, and a candidate EB in the Hyades that is solar-type with a likely substellar companion \citep[see also][]{mann2016}.  We also present updated models for the known EB Pleiades member HD 23642 using K2 data.

In Section~\ref{sec:data} we describe the data that we use, including the K2 light curves, photometry from the literature, and newly obtained spectroscopy. We describe our analysis procedures, including estimation of stellar properties and light-curve modeling, in Section~\ref{sec:analysis}. The results for the five EBs studied in this paper, including modeling results and initial physical parameters, are presented in Section~\ref{sec:results}. Finally, we briefly discuss the measured physical parameters in the context of stellar models in Section~\ref{sec:disc} and conclude with a summary in Section~\ref{sec:summary}.

\section{Data}
\label{sec:data}

As a part of the K2 Campaign 4 guest observer program (Fig.~\ref{fig:k2fov}), targets from \citet{stauffer2007} and \citet{sarro2014} were included in the proposed target list as long as they fell within reasonable brightness cuts. Known members of the Hyades were included from historical proper motion surveys \citep{vanbueren1952, vanaltena1969, hanson1975} as well as more recent surveys \citep{roser2011, goldman2013}.  

The $K2$ light curves for all Pleiades and Hyades members were examined by eye to identify potential EBs, and the membership of the detected EBs was then re-examined and confirmed using both archival and followup observations as described below.
The EB cluster members newly reported here and their ephemerides are summarized in Table~\ref{table:EBs}. Two previously known Pleiades EBs are summarized in Table~\ref{table:prevEBs}, one of which (HII 2407) was presented in detail in \citet{david2015a}.

\begin{deluxetable}{llllll}
\tabletypesize{\footnotesize}
\tablecolumns{6}
\tablecaption{Newly identified eclipsing binaries\label{table:EBs}}
\tablehead{
    \colhead{EPIC} & \colhead{Coordinates (J2000.0)} & \colhead{Common ID} & \colhead{Cluster} & \colhead{Period (d)} & \colhead{$BJD_0$ (BJD-2450000)} 
}
\startdata
210974364 & 03 42 27.30 +22 34 24.8 & HCG 76 & Pleiades & 32.747 & 7068.748 \\ 
211075914 & 03 46 55.31 +24 11 16.8 & MHO 9  & Pleiades &  42.8 & 7099.2 \\ 
210490365 & 04 13 05.60 +15 14 52.0 & vA 50 & Hyades & 3.48451 & 7062.5801 \\ 
\enddata
\end{deluxetable}

\begin{deluxetable}{llllll}
\tabletypesize{\footnotesize}
\tablecolumns{6}
\tablecaption{Previously known eclipsing binaries\label{table:prevEBs}}
\tablehead{
    \colhead{EPIC} & \colhead{Common ID} & \colhead{Cluster} & \colhead{Period (d)} & \colhead{$BJD_0$ (BJD-2450000)} & \colhead{Reference}
}
\startdata
211082420 & HD 23642 & Pleiades & $2.46113412 \pm 0.00000052$ & $7119.522069 \pm 0.00002$ & \cite{torres2003} \\
211093684 & HII 2407 & Pleiades & $7.0504829 \pm 0.0000047$ & $6916.65777 \pm 0.00014$ & \cite{david2015a} \\
\enddata
\end{deluxetable}

\subsection{K2 Photometry and Detrending}
Long-cadence ($\sim 30$ min exposure) \emph{Kepler} photometry was obtained for all requested targets.  Several different methods of data reduction and systematic removal were employed. Source photometry included the Simple Aperture Photometry (SAP) provided by the $Kepler$ project and available through MAST, as well as custom aperture photometry from the $Kepler$ target pixel files. The details of our custom aperture photometry procedure are discussed in \citet{david2016} and will be presented in detail in Cody et al. (2016, \emph{in prep}). Removal, or ``detrending'', of systematic trends related to jitter in the spacecraft pointing was achieved through the Gaussian process regression algorithm of \citet{aigrain2015}, the Pre-search Data Conditioning (PDC) procedure applied to the SAP flux, or a modified version of the Self-Flat-Fielding method \citep{vanderburg2014} which is described in detail in \citet{david2016}.

Our experience is that no one photometry method and no one detrending method can be considered best for all sources.  Thus our analysis makes use of the best available combination, chosen on a source-by-source basis. These decisions are based on an assessment of the photometric precision on 6.5 hour timescales (using the ``quasi-CDPP'' metric defined in \cite{aigrain2015}, the median value of the standard deviation in a moving window of a given duration) as well as visual inspection of the detrended light curves for the presence of remaining sawtooth-like systematic features related to spacecraft pointing.
In particular, 
for HCG 76 and MHO 9, the analyzed light curves were obtained from the \citet{aigrain2015} method of detrending applied to the SAP time series. For HD 23642, we used the PDC detrended SAP light curve, publicly available through MAST\footnote{Mikulski Archive for Space Telescopes - available at \url{http://archive.stsci.edu/index.html}}. Finally, for vA 50, we again used the PDC light curve, subject to additional detrending using the procedure described in \citet{david2016}.

\subsection{Photometric Colors\label{sec:colors}}

\begin{figure}[ht!]
\centering
\includegraphics[width=0.4\textwidth]{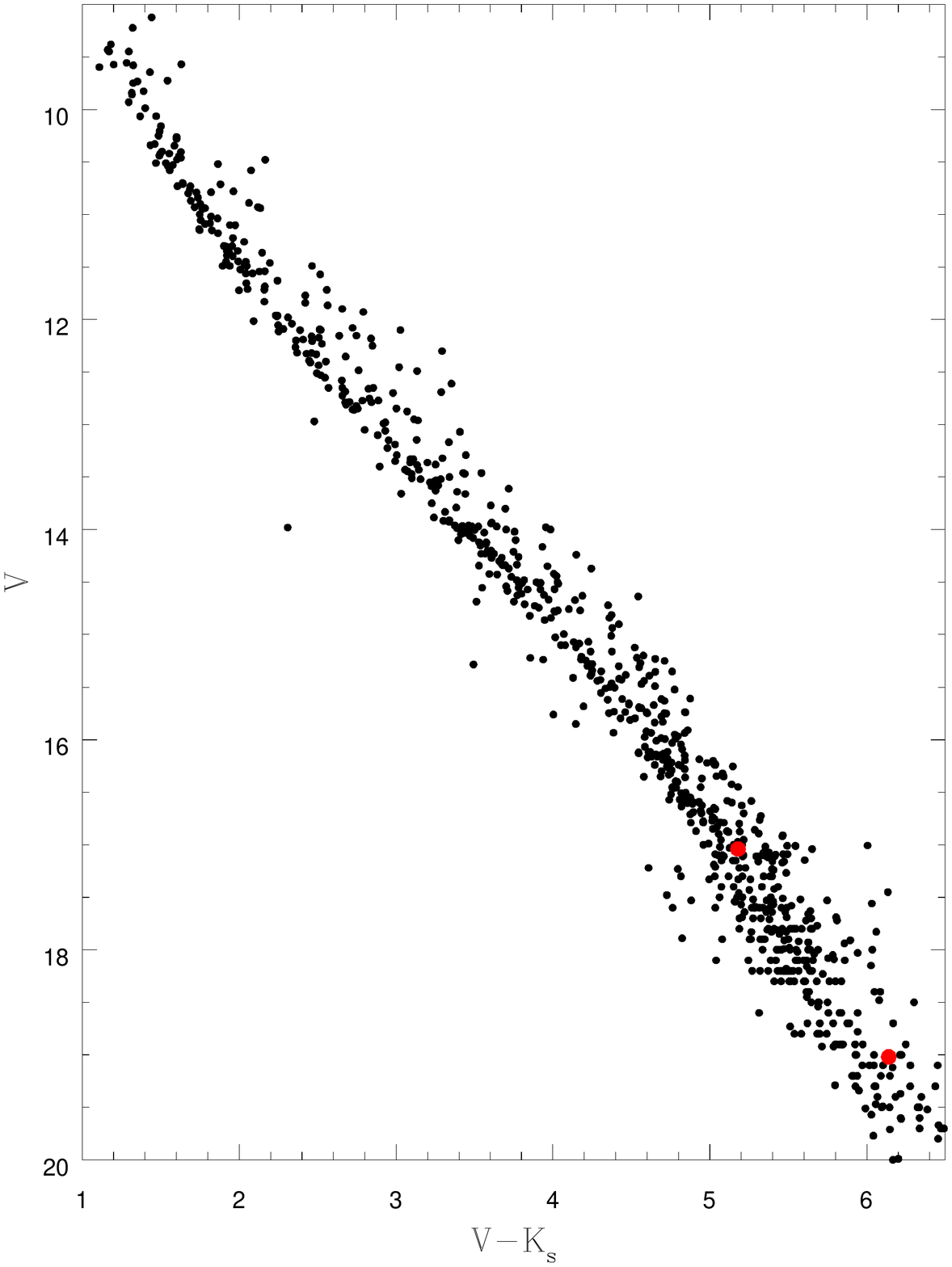}
\includegraphics[width=0.4\textwidth]{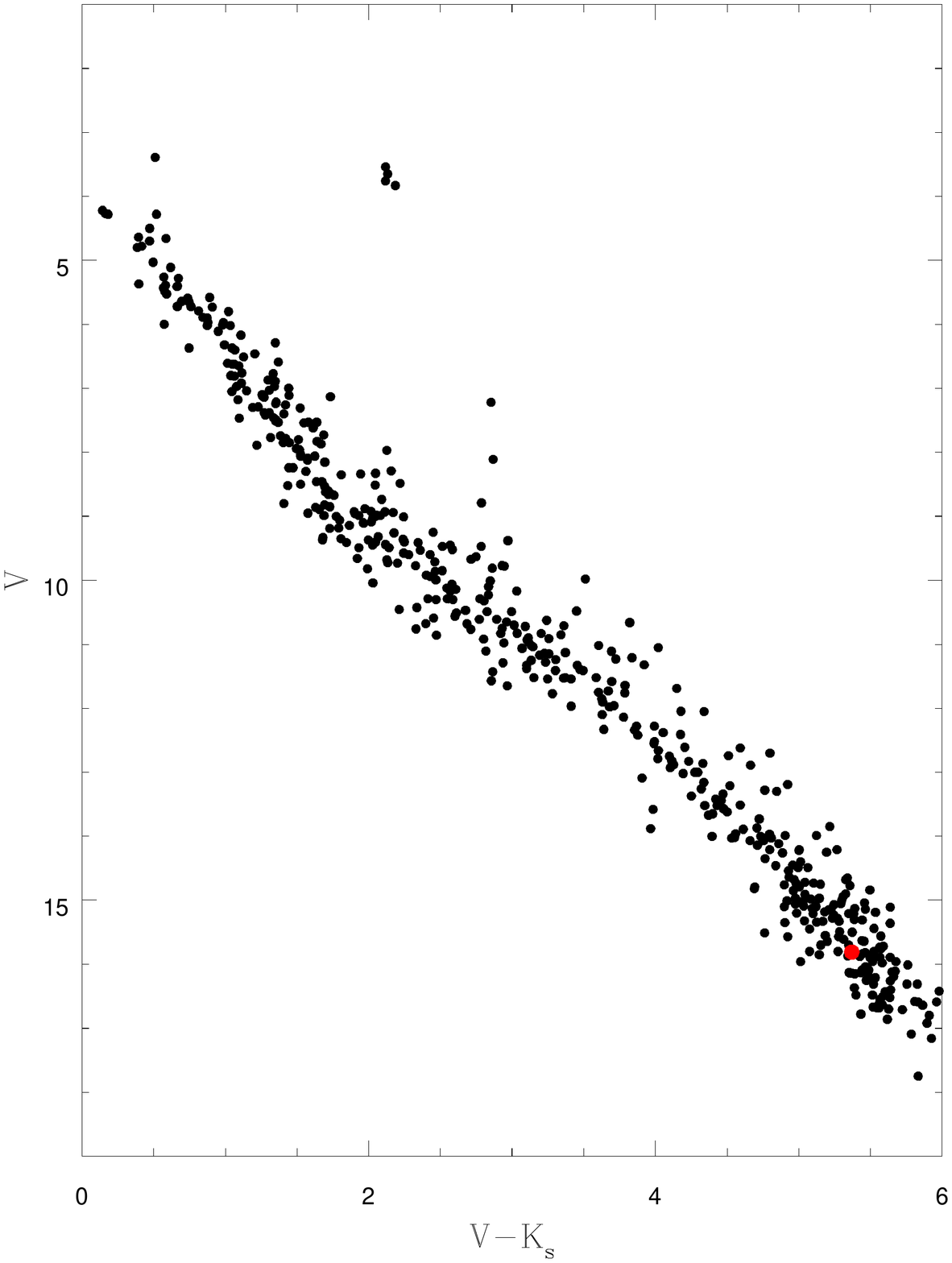}
\caption{$V$ vs $V-K_s$ photometric color magnitude diagram for the observed known members of the Pleiades (left) and Hyades (right) clusters.  The red highlighted points are the EBs reported in this paper.}
\label{fig:CMD}
\end{figure}

$V-K$ colors were assembled for each cluster target observed with $K2$.  The $K_s$ magnitudes are adopted from 2MASS \citep{cutri2012}.  The $V$ magnitudes are adopted from various sources for both the Pleiades \citep{stauffer1998, stauffer2007, kamai2014} and Hyades \citep{upgren1977, weis1979, weis1983, weis1985, weis1988}.   Figure~\ref{fig:CMD} shows color-magnitude diagrams for the Pleiades and Hyades, respectively, with our newly reported EBs highlighted, and Table \ref{table:EB_photometry} lists the adopted $V$ and $K_s$ magnitudes.

\begin{deluxetable}{lllllll}
\tabletypesize{\footnotesize}
\tablecolumns{6}
\tablecaption{Photometric magnitudes in $V$ and $K_s$ bands for reported EBs\label{table:EB_photometry}}
\tablehead{
    \colhead{EPIC} & \colhead{Common ID} & \colhead{Cluster} & \colhead{$V$} & \colhead{Reference} & \colhead{$K_s$} & \colhead{Reference}
}
\startdata
210974364 & HCG 76 & Pleiades & 17.04 & \citet{stauffer2007} & 11.86 & \citet{cutri2012} \\ 
211075914 & MHO 9  & Pleiades & 19.02 & \citet{stauffer1998} & 12.88 & \citet{cutri2012} \\ 
210490365 & vA 50 & Hyades & 15.81 & \citet{upgren1985} & 10.44 & \citet{cutri2012} \\
\enddata
\end{deluxetable}

\subsection{Spectroscopy}

Follow-up spectroscopy was obtained for all targets that were identified as potential new EB cluster members.  These spectra served to confirm membership in the cluster by verifying the systemic velocity as consistent with that of the cluster.  They also verify that the identified source is a spectroscopic binary, either right away for double-lined systems or following the acquisition of a time series for single-lined systems, and therefore exclude the possibility of a background EB contaminating the K2 light curve. Finally the spectra provide constraints on the physical properties of each system such as the primary spectral type and rotational velocity, and the primary (and secondary for double-lined systems) velocity amplitudes.  

Keck/HIRES \citep{vogt1994} spectra were collected at the epochs listed in
Table~\ref{table:rvs}. The images were processed and spectra extracted using either 
the California Planet Search (CPS) pipeline, requiring subsequent heliocentric correction, or the {\sc makee} package written by Tom Barlow. Radial velocities (RVs) were derived from Gaussian fitting to cross correlation
peaks using the routine \texttt{fxcor} within {\sc iraf}. Absolute calibration was achieved
for the M-type stars by baselining from Gl 176 and adopting the RV from \cite{nidever2002}; GJ 105B was used as a secondary standard.  
For the earlier type stars, a suite of GK-type reference stars from \cite{nidever2002} was used.
The derived radial velocities and measured flux ratios are listed in Table \ref{table:rvs}.

\begin{deluxetable}{cccrlrlc} 
\tabletypesize{\footnotesize} 
\tablewidth{\textwidth} 
\tablecaption{ Keck-I/HIRES Radial Velocities and Flux Ratios \label{table:rvs}} 
\tablehead{ 
\colhead{Proper} & \colhead{Epoch} & \colhead{Epoch} & \colhead{$v_1$} & \colhead{$\sigma_{v_1}$} & \colhead{$v_2$} & \colhead{$\sigma_{v_2}$} & \colhead{$F_2/F_1$} \\
\colhead{Name} & \colhead{(UT Date)} & \colhead{(BJD-2450000)} & \colhead{(\kms)} & \colhead{(\kms)} & \colhead{(\kms)} & \colhead{(\kms)} &
}
\startdata 
vA 50 &  20150925 & 7291.036879681 & 37.74  & 0.93    & \nodata & \nodata & $<$0.1  \\
\nodata   &	 20151003 & 7299.036548298 & 38.90  & 0.72    & \nodata & \nodata & $<$0.1  \\
\nodata   &  20151027 & 7322.880052281 & 37.60  & 1.21    & \nodata & \nodata & $<$0.1 \\
\nodata   &  20151031 & 7327.047595191 & 38.77  & 0.67    & \nodata & \nodata & $<$0.1 \\
\nodata	  &  20151113 & 7339.967314662 & 38.74  & 0.67    & \nodata & \nodata & $<$0.1 \\
\nodata	  &  20151128 & 7354.985788494 & 39.02  & 0.61    & \nodata & \nodata & $<$0.1 \\
\nodata	  &  20151129 & 7355.958388342 & 38.51  & 0.59    & \nodata & \nodata & $<$0.1 \\
HCG 76 & 20150925 & 7291.026244638 & -24.41  & 0.68  &  37.29  & 1.24 & 1.04 $\pm$ 0.20 \\
\nodata    & 20151001 & 7297.048107753 & -3.78   & 0.98  &  14.35  & 0.85 & 0.96 $\pm$ 0.08 \\
\nodata    & 20151003 & 7299.045895660 & 7.12    & 0.66$^\dagger$  &  \nodata   & \nodata & \nodata \\
\nodata    & 20151027 & 7322.865231574 & -23.31  & 0.48  & 35.76 & 0.27  & 1.04 $\pm$ 0.06 \\
\nodata    & 20151031 & 7327.039266404 & -18.94  & 0.69  & 31.42 & 0.89  & 0.95 $\pm$ 0.06 \\
\nodata    & 20151128 & 7354.992645023 & -20.65 & 0.69 & 35.37 & 0.66  & 0.94 $\pm$ 0.06 \\
\nodata    & 20151129 & 7355.949031859 & -23.24 & 0.73 & 37.07 & 0.63  & 0.88 $\pm$ 0.07 \\
\nodata    & 20151221 & 7377.834221745 & 19.73 & 0.37  & -10.39 & 0.37  & 0.92 $\pm$ 0.05 \\
\nodata    & 20151224 & 7380.724124841 & 4.67  & 0.44$^\dagger$  & \nodata & \nodata  & \nodata \\
\nodata    & 20151229 & 7385.918133654 & -14.21 & 0.47 &    27.51  &  0.43  &   0.96 $\pm$ 0.04 \\
MHO 9      & 20151027 & 7322.900583908 & -7.19 & 2.52 & 32.12 & 4.41 & 0.75 $\pm$ 0.17\\
\nodata    & 20151221 & 7377.788149023 & 10.23 & 1.83  & \nodata & \nodata & \nodata \\
\nodata    & 20151224 & 7380.878626538 & 12.64 & 2.21  & \nodata & \nodata & \nodata \\
\nodata    & 20151229 & 7385.786528022 & 14.72 & 2.05  & -21.85  & 5.13 & 0.39 $\pm$ 0.12 \\
\nodata   & 20160124 & 7411.775413410 & -0.67 & 1.63 & \nodata & \nodata & \nodata \\
AK II 465   & 20151027 & 7323.164355309 &  57.18      & 0.59 & 0.25  & 0.52           &  0.76 $\pm$ 0.09 \\
\nodata    & 20151221 & 7377.933238088 & -27.34 & 0.84 & 73.68 & 0.57 & 0.75 $\pm$ 0.12 \\
\nodata    & 20151224 & 7381.023604991 & 97.07     & 0.72   &  -35.90     & 0.38      & 0.70 $\pm$ 0.10             \\
\nodata    & 20151229 & 7385.935177983 &  -28.34     &  0.53     & 77.19     & 0.52   & 0.76 $\pm$ 0.11 \\
HD 23642$^*$  & 20151224 & 7381.026716140 & -90.44 & 4.58 & 147.75 & 3.30 & \nodata         
\enddata 
\tablecomments{Quoted radial velocities are weighted means across several spectral orders within a single epoch, with each measurement weighted inversely to the variance. The uncertainties used in the orbital parameter fitting procedure are the root-mean-square errors between individual measurements. The final column lists flux ratios, measured from the relative peak heights in the cross-correlation functions. $^\dagger$In the orbit fitting of HCG 76 we used an {\it ad hoc} uncertainty of 3 \kms\ for the 20151003 and 20151224 epochs, believing the formal values in the table to be an underestimate due to the small velocity separation between components relative to the spectrograph resolution. $^*$Though we report a single epoch of RVs here for HD 23642 we did not include these measurements in the analysis that follows. The data are consistent with our solution, however.
}
\end{deluxetable}

\section{Analysis}
\label{sec:analysis}

Our analysis consists of measuring or estimating properties of the primary star in each EB, 
and fitting the combined photometric and RV time-series data in order to 
derive the properties of the secondaries. Here we briefly describe some of the general
analysis methods that were used in common to all EB systems. Individualized analysis for 
specific EBs appears below.

\subsection{Estimation of Primary Star Properties}\label{sec:stellarprops}

We determine the absolute $V$ and $K_s$ magnitudes from the measured or adopted apparent magnitudes and colors (see Section \ref{sec:colors}) using bolometric corrections from \citet{pecaut2013}, assuming a distance of 136.2 pc for the Pleiades \citep{melis2014} and $\sim 45$ pc for the Hyades \citep{perryman1998, debruijne2001}.
Effective temperatures are estimated from the adopted $V-K_s$ colors (see Section~\ref{sec:colors}) using the following relations, derived empirically by fitting polynomials to color and temperature data from \citet{pecaut2013}, valid for $0.3 < V-K_s < 7.0$:
\begin{equation}
T_\mathrm{eff} [K] = \frac{5000.0}{0.51903 + 0.24918 (V-K_s) - 0.02160 (V-K_s)^2 + 0.00415 (V-K_s)^3 - 0.000359 (V-K_s)^4}
\end{equation}
\noindent
Radii of the primary stars can then be estimated using the Stefan-Boltzman law, adopting $T_{\mathrm{eff},\odot} = 5770$~K.

Masses of the primary stars can be estimated from empirical relations. For the lowest mass stars with $V-K > 4.0$, we adopt the
relation derived by \citet{delfosse2000} for $4.0 < V-K < 7.0$:
\begin{equation}
\begin{aligned}
\log (M/M_\odot) = {} & 0.001 \times [7.4 + 17.61 (V-K_s) +33.216 (V-K_s)^2 + 34.222 (V-K_s)^3  \\ 
& -27.1986 (V-K_s)^4 +4.94747 (V-K_s)^5 -0.27454 (V-K_s)^6 ]
\end{aligned}
\end{equation}

Note that these estimates are approximations only.  These empirical relations are nominal for main sequence stars with solar metallicity.  The Hyades is slightly metal rich and the Pleiades, although essentially at solar metallicity, is still pre-main sequence at the lowest stellar masses.

\subsection{Light Curve Modeling}
\label{subsec:modeling}

Modeling of the EB light curves was performed with the publicly available code \jktebop \footnote{\url{http://www.astro.keele.ac.uk/jkt/codes/jktebop.html}} \citep[and references therein]{southworth2013}.
\jktebop\ is based on the Eclipsing Binary Orbit Program \citep{popper1981, etzel1981}, which relies on the Nelson-Davis-Etzel biaxial ellipsoidal model for well-detached EBs \citep{nd1972, etzel1975}. \jktebop\ models the two components as biaxial spheroids for the calculation of the reflection and ellipsoidal effects, and as spheres for the eclipse shapes.  \jktebop\ finds the best-fit model to a light curve through Levenberg-Marquardt (L-M) optimization. Robust statistical errors on the best-fit model parameters are then found through repeated Monte Carlo (MC) simulations in which Gaussian white noise commensurate to the observational errors is added to the best-fit model. A new L-M fit is performed on the perturbed best-fit model and the new parameters are saved as links in the MC chain. The final orbital parameters for each system are then given by the original L-M best-fit, with uncertainties given by the standard deviations determined from the MC parameter distributions.  

All modeling in this work took into account the effect of the $\approx$30~min \emph{Kepler} K2 cadence by numerically integrating model light curves at ten points in a total time interval of 1766 seconds, corresponding to the integration time of Kepler long cadence data.

\section{Results}
\label{sec:results}


\subsection{HCG 76}
\label{sec:hcg76}

HCG 76 (V 612 Tau, EPIC 210974364) was first identified as a probable Pleiades member, and given the HCG designation, when it was detected as a flare star \citep{haro1982}.  In a proper motion membership study of HCG stars, \citet{stauffer1991} found HCG 76 to have a membership probability of 90\%.  Subsequent proper motion surveys of the Pleiades re-identified HCG 76 as a member and provided the alternate designations of HHJ 294 and DH 224 \citep{hambly1993, deacon2004}. The spectral type estimate based on colors is $\sim$M3.

\begin{figure}[ht!]
\centering
\includegraphics[width=0.99\textwidth]{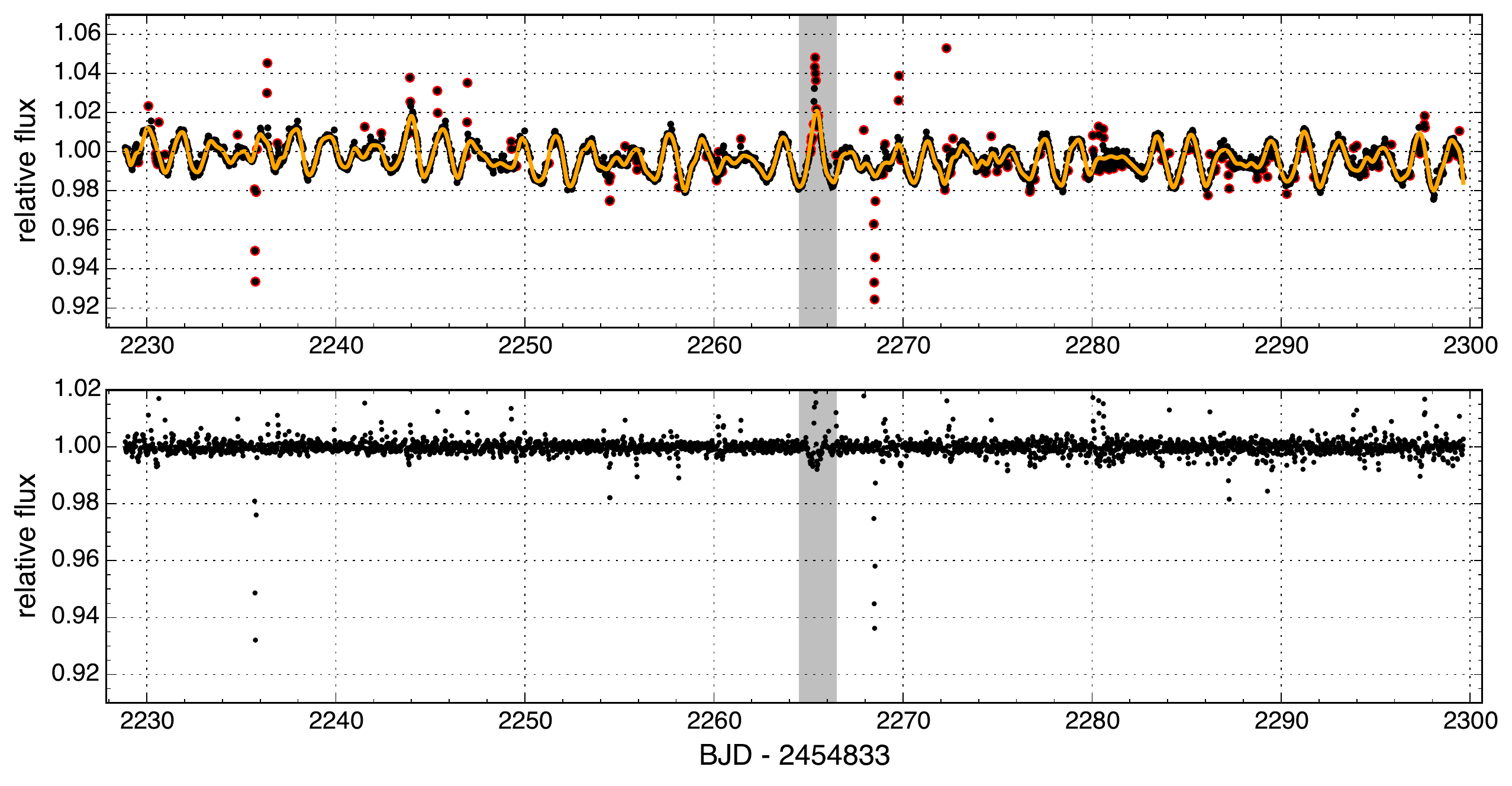}
\caption{Top panel: The systematics corrected $K2$ light curve for HCG 76 with our variability fit indicated by the orange curve. Outlier points excluded from this fit are marked by the red circles. Bottom panel: The rectified light curve, from dividing out the variability fit above, upon which we performed our fitting procedure. In both panels the gray shaded region highlights a portion of the light curve that is poorly modeled by the variability fit, leading to the introduction of systematics in the rectified light curve.}
\label{fig:hcg76varfit}
\end{figure}

\begin{figure}[ht!]
\centering
\includegraphics[width=0.75\textwidth]{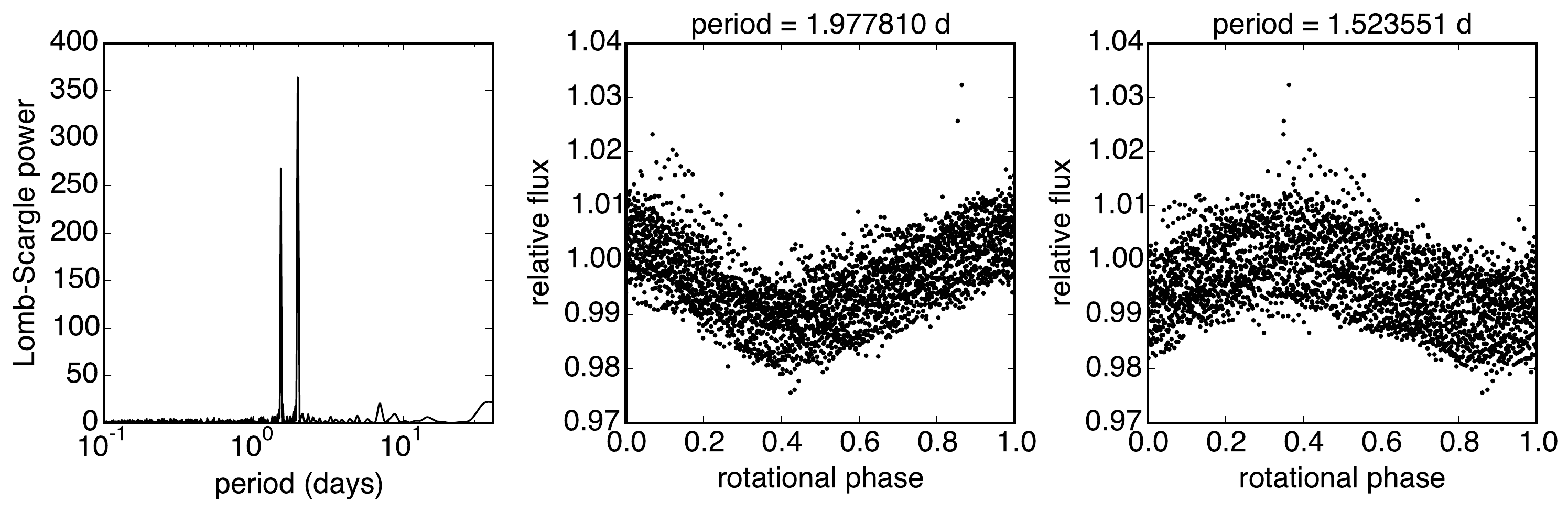}
\caption{Lomb-Scargle periodogram of the variable light curve for HCG 76 (left) and the K2 light curve for the object phase folded on the two strong rotational periods detected in the periodogram (middle and right panels). Outliers (both flares and eclipses) have been removed from the light curves in these figures for the purposes of illustrating the sinusoidal rotation signals.}
\label{fig:hcg76-rotation}
\end{figure}

The $K2$ light curve (Fig.~\ref{fig:hcg76varfit}) used in our analysis was corrected for systematics from the raw SAP flux using the algorithm described in \cite{aigrain2015}. The light curve is characterized by a beating spot pattern, likely due to the different rotation periods of the primary and secondary. A Lomb-Scargle periodogram analysis on the systematics-corrected photometry identifies significant periodicity at 1.524$\pm$0.028~d and 1.978$\pm$0.051~d (see Fig.~\ref{fig:hcg76-rotation}). In each case, the rotational period uncertainty has been coarsely approximated from the full width at half maximum (FWHM) of the corresponding periodogram peak. The variability was then iteratively fit with a cubic B-spline, using breakpoints every 12 cadences or $\sim$6~hr, and 2-$\sigma$ low or 4-$\sigma$ high outlier exclusion upon each iteration, using the method described in \cite{david2016}.

The light curve shows two primary eclipses with depth $\sim$7\%, plus two secondary eclipses with depth $\sim$2\% slightly offset from phase = 0.5.  The eclipses are of short duration, with only a few points in eclipse.  Due to these factors, a periodogram analysis fails; the initial period estimate by eye is $\approx$33~d and that from the orbital fit below is 32.7~d. 

\begin{figure}[ht!]
\centering
\includegraphics[width=0.75\textwidth]{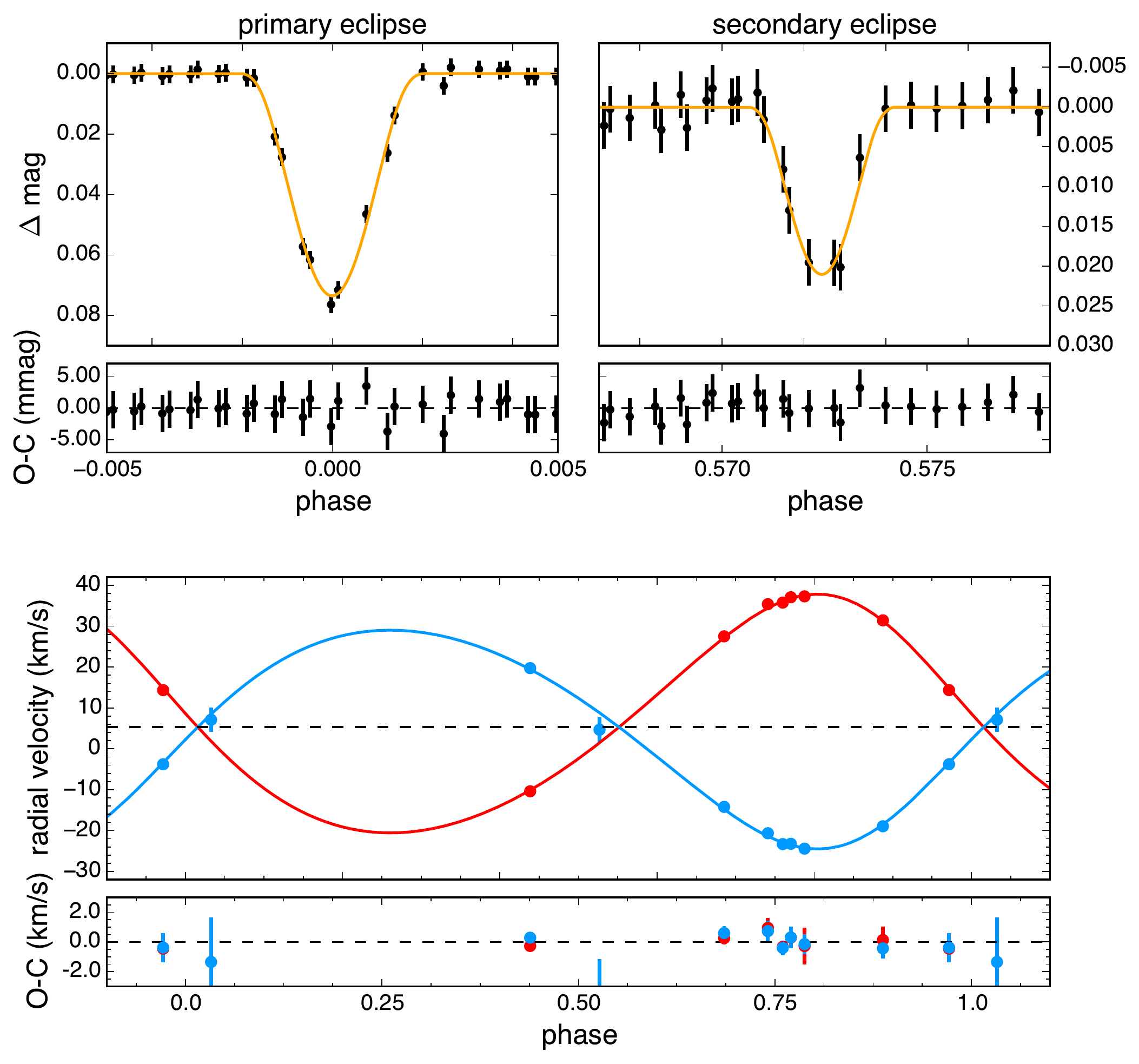}
\caption{Best-fit \jktebop\ model to the $K2$ photometry (top) and Keck/HIRES RVs (bottom) for HCG 76.}
\label{fig:hcg76fit}
\end{figure}

The corrected K2 light curve and follow-up Keck/HIRES RVs (Table \ref{table:rvs}) were used to determine a best-fit \jktebop\ model of the system. We assumed a linear limb darkening law with coefficient $u$=0.6 for each component, though because the eclipses are only partial and the ingress and egress are sparsely sampled, limb darkening cannot be strongly constrained with current data. Table \ref{table:hcg76table} lists the best-fit parameters from the model shown in Fig.~\ref{fig:hcg76fit}. The uncertainty in the ratio of the radii is large, despite the spectroscopic light ratio constraints imposed, due to the fact that the eclipses are grazing and $e$ and $\omega$ are currently poorly  constrained. Additional RVs covering the first half of the orbit, as well as high cadence photometry of the eclipses to fill out the light curve, will help to further constrain the masses and radii of this system.

The Monte Carlo distributions in mass and radius for the binary components include solutions for which the system is highly non-coeval, with the less massive component falling below the main sequence of solar metallicity BHAC15 models. This tail of physically implausible solutions could be cleanly separated from the densely populated and more likely region of parameter space by considering only those solutions with $R_1/R_2<1.4$. Excluding the implausible solutions, we obtain $\lesssim$4.5\% precision in the component radii. Specifically, we find $M_1=0.3019 \pm 0.0070$ \msun, $M_2=0.2767 \pm 0.0068$ \msun, $R_1=0.341 \pm  0.016$ \rsun, and $R_2 = 0.319 \pm 0.013$ \rsun. In Table \ref{table:hcg76table}, we thus report all relevant parameter values both including and excluding these implausible solutions. Notably, the dynamical masses are $\sim$30\% larger than the value predicted by the combined light $V-K_s$ color and the \cite{delfosse2000} relation presented in \S~\ref{sec:stellarprops}. The mass-$K_s$ relation for M-dwarfs presented in \cite{mann2015} is slightly more consistent with our dynamical masses, predicting a mass that is $\sim$10-20\% smaller than our fundamental values, after correcting the absolute $K_s$ magnitude for binarity. We note that the \cite{mann2015} relation was derived for main-sequence stars, while HCG 76 is still pre-main-sequence and thus more luminous compared to their MS counterparts.

\begin{figure}[ht!]
\centering

\includegraphics[width=0.99\textwidth]{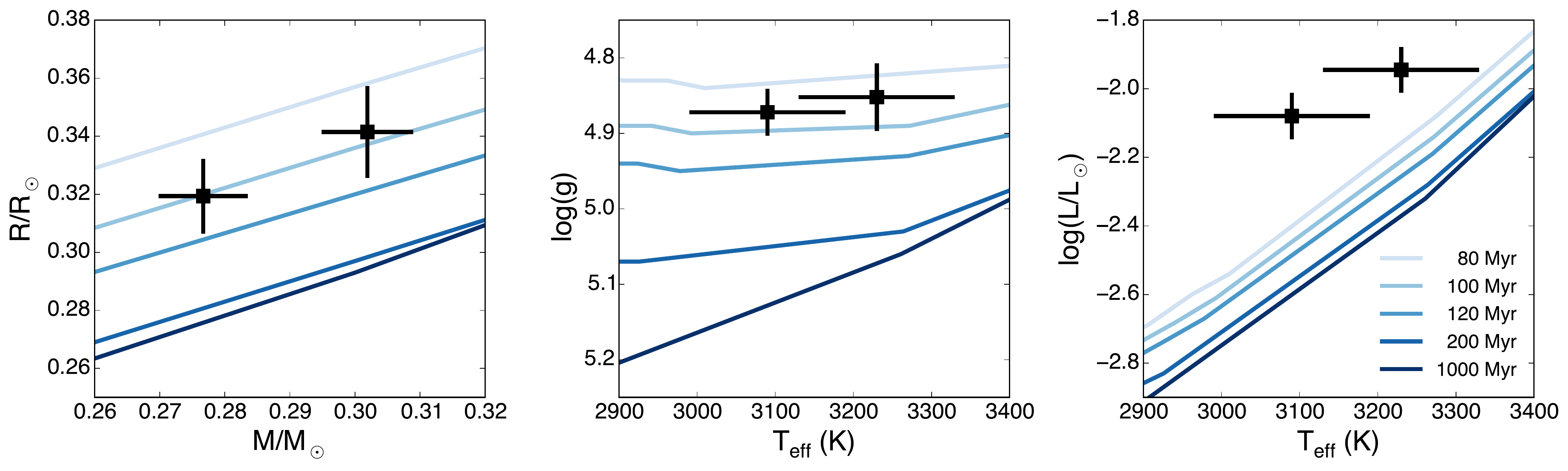}
\caption{The positions of HCG 76 components relative to BHAC15 isochrones in the mass-radius plane (left), $T_\mathrm{eff}-\log{g}$ plane (middle), and $T_\mathrm{eff}-\log{L/L_\odot}$ plane (right). Square points represent best-fit values and errorbars indicate 1-$\sigma$ uncertainties. We assumed 100 K uncertainties in the temperatures and propagated these through in determining the luminosity uncertainties. The two components are consistent within error of being coeval in the mass-radius plane at $\sim$100 Myr, though they appear younger in the $T_\mathrm{eff}-\log{g}$ plane. The luminosities calculated from the Stefann-Boltzmann law, the measured radii, and photometric temperatures, are significantly larger than the model predictions.}
\label{fig:hcg76bhac15}
\end{figure}

\begin{figure}
    \centering
    \includegraphics[width=0.99\textwidth]{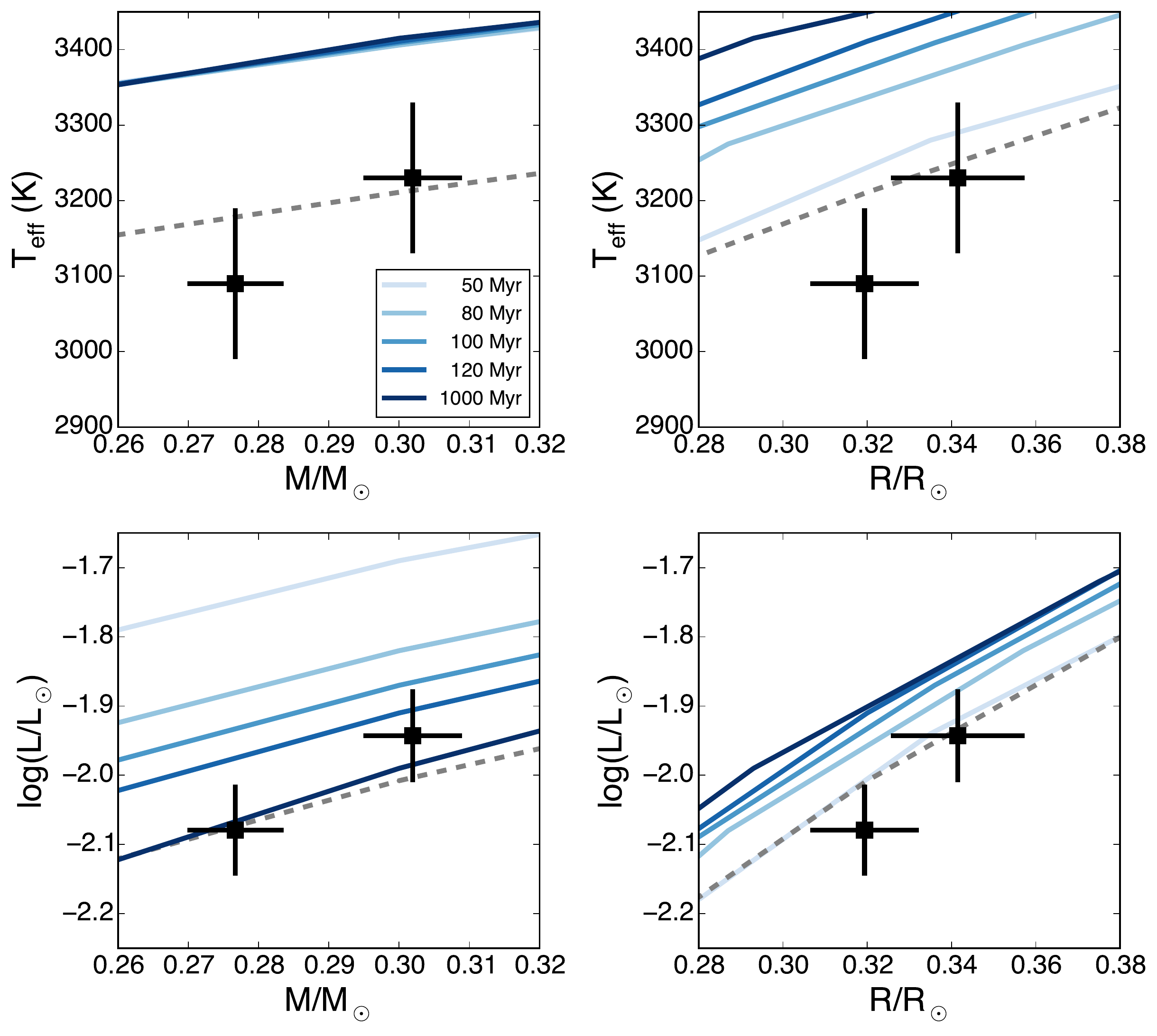}
    \caption{BHAC15 isochrones in the mass-temperature, radius-temperature, mass-luminosity, and radius-luminosity planes (clockwise from upper left panel) compared to the derived parameters for the HCG 76 components. In each panel, the dashed line indicates the 120 Myr isochrone shifted by 200 K towards cooler effective temperatures (or the luminosities resulting from such a shift).}
    \label{fig:hcg76-4panel}
\end{figure}

In Figures \ref{fig:hcg76bhac15} and \ref{fig:hcg76-4panel}we compare the \jktebop\ derived parameters for this system with solar metallicity ($Z$=0.02) \cite{baraffe2015} evolutionary models, hereafter BHAC15. Figure \ref{fig:hcg76bhac15} shows that the components are similar in mass and are both consistent with the slope of the isochrones in the mass-radius and temperature-gravity planes, but preferring a slightly younger age than 120 Myr as suggested by \citet{dahm2015}. 
The component ages derived using the \texttt{griddata} routine in {\sc Python} 
are $\tau_1$=106$^{+18}_{-13}$ Myr and $\tau_2$=102$^{+24}_{-22}$ Myr, where the median and 68\% confidence intervals are quoted. The two stars are thus consistent with being coeval within the uncertainties. A more precise system age can be determined using the assumption of coevality from the product of the two age distribution functions. The resulting system age is $\tau$=103$^{+7}_{-10}$ Myr.
If we again assume coevality and age-date the system in the $T_\mathrm{eff}$-$\log{g}$ plane, 
we find a mode age of 93 Myr, with a median and 68\% confidence interval of 102$^{+112}_{-14}$ Myr. The long tail towards older ages is due to the clustering of isochrones towards the main sequence.

Figure~\ref{fig:hcg76-4panel} illustrates inconsistencies between the observations and the BHAC15 models; specifically, at a fixed age, the models are unable to simultaneously reproduce mass, radius, \teff, and luminosity. We suggest that these discrepancies can be largely resolved, at least in this narrow mass range, by shifting the models by 200 K towards cooler temperatures. It is possible that the models are incorrectly predicting \teff, or that there is a systematic offset in the adopted empirical color-\teff\ or spectral type-\teff\ conversions, or some combination of both effects.  Such a temperature shift would likewise improve the agreement among the panels in Figure \ref{fig:hcg76bhac15}.

Finally, we can use the highly precise stellar parameters that we have determined for HCG~76 to make an independent measurement of the distance to the Pleiades. We used the broadband catalog photometry assembled by \citet{sarro2014} and supplemented these with the available WISE photometry. In total, the observed broadband spectral energy distribution (SED) spans a wavelength range of 0.35--12 $\mu$m. The SED was fit by the sum of two NextGen stellar atmosphere models \citep{Hauschildt1999} of solar metallicity, interpolated to the respective HCG~76 component \teff\ and $\log g$, and scaled by the respective radii squared. We adopted the canonical Pleiades extinction of $A_V = 0.12$. We varied the component \teff's and radii within their uncertainties (Table~\ref{table:hcg76table}) but enforcing the directly measured \teff\ ratio and sum of radii. The observed $u$-band flux exhibits a strong excess over the nominal SED, not surprising considering the identification of HCG~76 as an active flaring star \citep{Kazarovets1993}. Excluding the $u$-band flux, the resulting SED fit is excellent, with reduced $\chi^2$ of 2.4. The corresponding distance is 132$\pm$5 pc, consistent with most recent determinations of the Pleiades distance \citep[see Table 8 in][]{southworth2005} including their results on the massive Pleiades EB HD~23642 that is also discussed below, 
and the VLBI distance to a different Pleiades member of 136.2$\pm$1.2 pc \citep{melis2014}.

\newcommand{\period}{32.7470}
\newcommand{\eperiod}{0.0013}
\newcommand{\tzero}{68.7480}
\newcommand{\etzero}{0.0010}
\newcommand{\sbratio}{0.84}
\newcommand{\esbratio}{0.34}
\newcommand{\sumradii}{0.0184}
\newcommand{\esumradii}{0.0022}
\newcommand{\ratradii}{0.93}
\newcommand{\eratradii}{0.23}
\newcommand{\inclination}{89.13}
\newcommand{\einclination}{0.17}
\newcommand{\ecosw}{0.1138}
\newcommand{\eecosw}{0.0040}
\newcommand{\esinw}{0.0682}
\newcommand{\eesinw}{0.0087}
\newcommand{\rvprim}{29.19}
\newcommand{\ervprim}{0.30}
\newcommand{\rvsec}{26.75}
\newcommand{\ervsec}{0.31}
\newcommand{\rvsys}{5.31}
\newcommand{\ervsys}{0.17}
\newcommand{\rada}{0.00890}
\newcommand{\erada}{0.00099}
\newcommand{\radb}{0.0095}
\newcommand{\eradb}{0.0030}
\newcommand{\lightratio}{0.950}
\newcommand{\elightratio}{0.022}
\newcommand{\ecc}{0.1326}
\newcommand{\eecc}{0.0049}
\newcommand{\omegap}{30.9}
\newcommand{\eomegap}{3.7}
\newcommand{\bpri}{1.57}
\newcommand{\ebpri}{0.58}
\newcommand{\bsec}{1.80}
\newcommand{\ebsec}{0.65}
\newcommand{\rchisq}{0.412}
\newcommand{\erchisq}{0.799}
\newcommand{\rmslc}{1.627}
\newcommand{\rmsprv}{0.438}
\newcommand{\rmssrv}{1.439}
\newcommand{\sma}{35.88} 
\newcommand{\esma}{0.29} 
\newcommand{\mrat}{0.917} 
\newcommand{\emrat}{0.013} 
\newcommand{\mpri}{0.2768} 
\newcommand{\empri}{0.0072} 
\newcommand{\msec}{0.3020} 
\newcommand{\emsec}{0.0073} 
\newcommand{\rpri}{0.319} 
\newcommand{\erpri}{0.036} 
\newcommand{\rsec}{0.34} 
\newcommand{\ersec}{0.11} 
\newcommand{\logga}{4.87} 
\newcommand{\elogga}{0.11} 
\newcommand{\loggb}{4.85} 
\newcommand{\eloggb}{0.22} 
\newcommand{\rhoa}{8.5} 
\newcommand{\erhoa}{4.4} 
\newcommand{\rhob}{7.7} 
\newcommand{\erhob}{3.2} 

\begin{deluxetable}{lrlll}
\tabletypesize{\scriptsize}
\tablecaption{System Parameters of HCG 76 \label{table:hcg76table}}
\tablewidth{0.99\textwidth}
\tablehead{
\colhead{Parameter} & \colhead{Symbol} & \colhead{JKTEBOP} & \colhead{Adopted} & \colhead{Units}
\\
& & \colhead{Value} & \colhead{Value} &
} 

\startdata
Orbital period & $P$ & \period\ $\pm$ \eperiod & 32.7470 $\pm$ 0.0013 & days \\
Ephemeris timebase - 2457000 & $T_0$ & \tzero\ $\pm$ \etzero & 68.748 $\pm$ 0.0010 & BJD \\
Surface brightness ratio & $J$ & \sbratio\ $\pm$ \esbratio & 0.84 $\pm$ 0.12 & \\
Sum of fractional radii & $(R_1+R_2)/a$ & \sumradii\ $\pm$ \esumradii & 0.01842 $\pm$ 0.00050 & \\
Ratio of radii & $k$ & \ratradii\ $\pm$ \eratradii & 0.938 $\pm$ 0.065 &  \\

Orbital inclination & $i$ & \inclination\ $\pm$ \einclination & 89.126 $\pm$ 0.029 & deg \\
Combined eccentricity, periastron longitude & $e\cos\omega$ & \ecosw\ $\pm$ \eecosw & 0.1137 $\pm$ 0.0011 & \\
Combined eccentricity, periastron longitude & $e\sin\omega$ & \esinw\ $\pm$ \eesinw & 0.0681 $\pm$ 0.0085 & \\
Primary radial velocity amplitude & $K_1$ & \rvsec\ $\pm$ \ervsec & 26.75 $\pm$ 0.30 & km s$^{-1}$ \\
Secondary radial velocity amplitude & $K_2$ & \rvprim\ $\pm$ \ervprim & 29.19 $\pm$ 0.29 & km s$^{-1}$ \\
Systemic radial velocity & $\gamma$ & \rvsys\ $\pm$ \ervsys & 5.31 $\pm$ 0.17 & km s$^{-1}$ \\
Fractional radius of primary & $R_1/a$ & \radb $\pm$ \eradb & 0.00952 $\pm$ 0.00047 & \\
Fractional radius of secondary & $R_2/a$ & \rada $\pm$ \erada & 0.00890 $\pm$ 0.00033 & \\
Luminosity ratio & $L_2/L_1$ & \lightratio\ $\pm$ \elightratio & 0.951 $\pm$ 0.022 & \\
Eccentricity & $e$ & \ecc\ $\pm$ \eecc & 0.1328 $\pm$ 0.0043 & \\
Periastron longitude & $\omega$ & \omegap\ $\pm$ \eomegap & 30.8 $\pm$ 3.2 & deg \\
Impact parameter of primary eclipse & $b_1$ & \bsec\ $\pm$ \ebsec & 1.807 $\pm$ 0.068 & \\
Impact parameter of secondary eclipse & $b_2$ & \bpri\ $\pm$ \ebpri & 1.577 $\pm$ 0.077 & \\
Orbital semi-major axis & $a$ & \sma\ $\pm$ \esma & 35.88 $\pm$ 0.27 & \rsun  \\
Mass ratio & $q$ & \mrat\ $\pm$ \emrat & 0.917 $\pm$ 0.013 & \\
Primary mass & $M_1$ & \msec\ $\pm$ \emsec & 0.3019 $\pm$ 0.0070 & \msun \\
Secondary mass & $M_2$ & \mpri\ $\pm$ \empri &  0.2767 $\pm$ 0.0068 & \msun \\
Primary radius & $R_1$ & \rsec\ $\pm$ \ersec & 0.341 $\pm$ 0.016 & \rsun \\
Secondary radius & $R_2$ & \rpri\ $\pm$ \erpri & 0.319 $\pm$ 0.013 & \rsun \\
Primary surface gravity & $\log g_1$ & \loggb\ $\pm$ \eloggb & 4.852 $\pm$ 0.045 & cgs \\
Secondary surface gravity & $\log g_2$ & \logga\ $\pm$ \elogga & 4.872 $\pm$ 0.031 & cgs \\
Primary mean density & $\rho_1$ & \rhob\ $\pm$ \erhob & 7.7 $\pm$ 1.3 & $\rho_\odot$ \\
Secondary mean density & $\rho_2$ & \rhoa\ $\pm$ \erhoa & 8.56 $\pm$ 0.95 & $\rho_\odot$ \\
Temperature ratio & $T_2/T_1$ & 0.957 $\pm$ 0.099 & 0.957 $\pm$ 0.034 & \\
Primary temperature & $T_1$ & \nodata & 3230 $\pm$ 100 &  K\\
Secondary temperature & $T_2$ & \nodata & 3090 $\pm$ 100  & K\\
Reduced chi-squared of light curve fit & $\chi^2_\mathrm{red}$ & \rchisq & \nodata & \\
RMS of best fit light curve residuals & & \rmslc & \nodata & mmag \\
Reduced chi-squared of primary RV fit & $\chi^2_\mathrm{red}$ & 0.007 & \nodata & \\
RMS of primary RV residuals & & \rmsprv & & km s$^{-1}$ \\
Reduced chi-squared of secondary RV fit & $\chi^2_\mathrm{red}$ & 0.025 & \nodata & \\
RMS of secondary RV residuals & & \rmssrv & \nodata & km s$^{-1}$
\tablecomments{The \jktebop\ best-fit orbital parameters and 1-$\sigma$ uncertainties result from 5,000 Monte Carlo simulations. The primary temperature is calculated from the $V-K_s$ color, as described in \S~\ref{sec:stellarprops}. The adopted value column indicates the mean and 1-$\sigma$ errors of the Monte Carlo parameter distributions after excluding those physically implausible solutions with $R_1/R_2>$1.4. See \S~\ref{sec:hcg76} for details.}
\enddata

\end{deluxetable}


\subsection{MHO 9 (BPL 116)}

MHO 9 (BPL 116, EPIC 211075914) was first identified as a candidate Pleiades member of very low mass in \citet{stauffer1998}, based on $V$ and $I$ photometry obtained with the Mt. Hopkins 48" telescope, and on proper motion consistent with the Pleiades derived from UK Schmidt plates\footnote{We caution the reader that there is another star in Taurus with the designation MHO 9, not to be confused with the Pleiades EB discussed here. Identified in \cite{briceno1998}, that star is a pre-MS weak-lined T-Tauri star, also of moderately late M type.}.  An independent combined photometric ($IZ$) and proper motion survey also identified it as a probable very low mass Pleiades member, under the designation BPL 116 \citep{pinfield2000}. The star was later confirmed as a Pleiades proper motion member by \citet{deacon2004}.   The spectral type estimate based on colors is $\sim$M4.5.

\begin{figure}[ht!]
\centering
\includegraphics[width=0.75\linewidth]{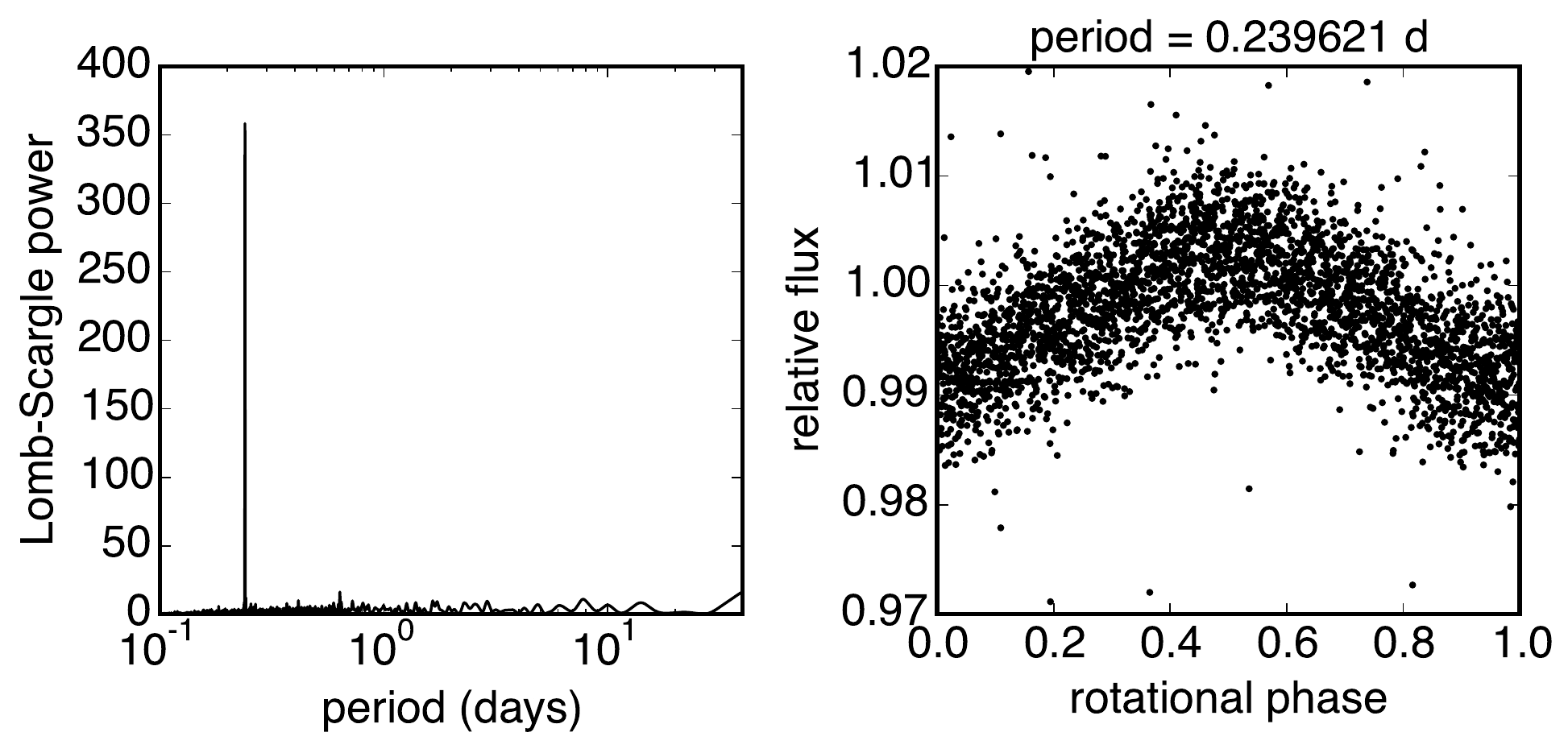}
\caption{Lomb-Scargle periodogram for the systematics corrected light curve of MHO 9 (left), phased at the inferred rotation period (right). Eclipses are excluded in the scaling of this figure for clarity.} 
\label{fig:bpl116rot}
\end{figure}

The detrended K2 light curve exhibits periodic, low amplitude undulations, which we interpret as due to spot rotation with a single period of $P_\mathrm{rot}$ = 0.2396 $\pm$ 0.0008 d (see Fig.~\ref{fig:bpl116rot}). The uncertainty in $P_{\rm rot}$ has been coarsely approximated from the FWHM of the Lomb-Scargle periodogram peak.

The Keck/HIRES spectrum reveals a double-lined system with the individual RVs reported in Table~\ref{table:rvs}.  The rotational velocity can be estimated by broadening standard star templates that are then added together with the measured RV difference and the inferred flux ratio from the cross correlation analysis.  The resulting estimate of $v\sin{i} \approx 42 \pm 1$ \kms\ assumes that the two components of the binary have the same projected velocity, as the lines are close enough together that it is not possible to fit separately for each component. Combining this $v\sin{i}$ estimate with our $P_{\rm rot}$ estimate above, and assuming the rotational axis of the primary is perpendicular to our line-of-sight, we arrive at an approximation of $R_1 \approx 0.199 \pm 0.005$ \rsun\ for the primary radius, where the uncertainty is the formal statistical error.

\begin{figure}[ht!]
\centering
\includegraphics[width=0.75\linewidth]{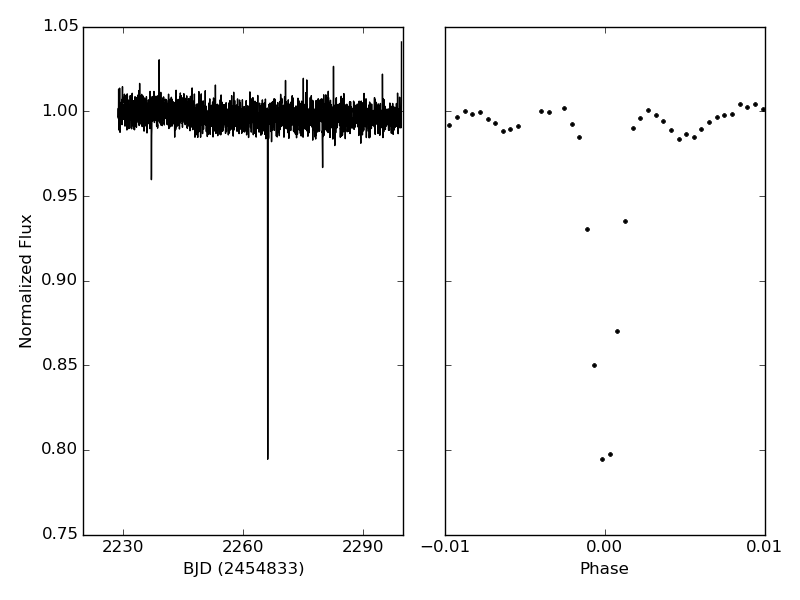}
\caption{Detrended normalized light curve for MHO 9 (BPL 116).  The full unphased light curve is shown on the left with the phased light curve zoomed in on the primary eclipse shown on the right.}
\label{fig:bpl116dtr}
\end{figure}

\begin{figure}[ht!]
\centering
\includegraphics[width=0.7\linewidth]{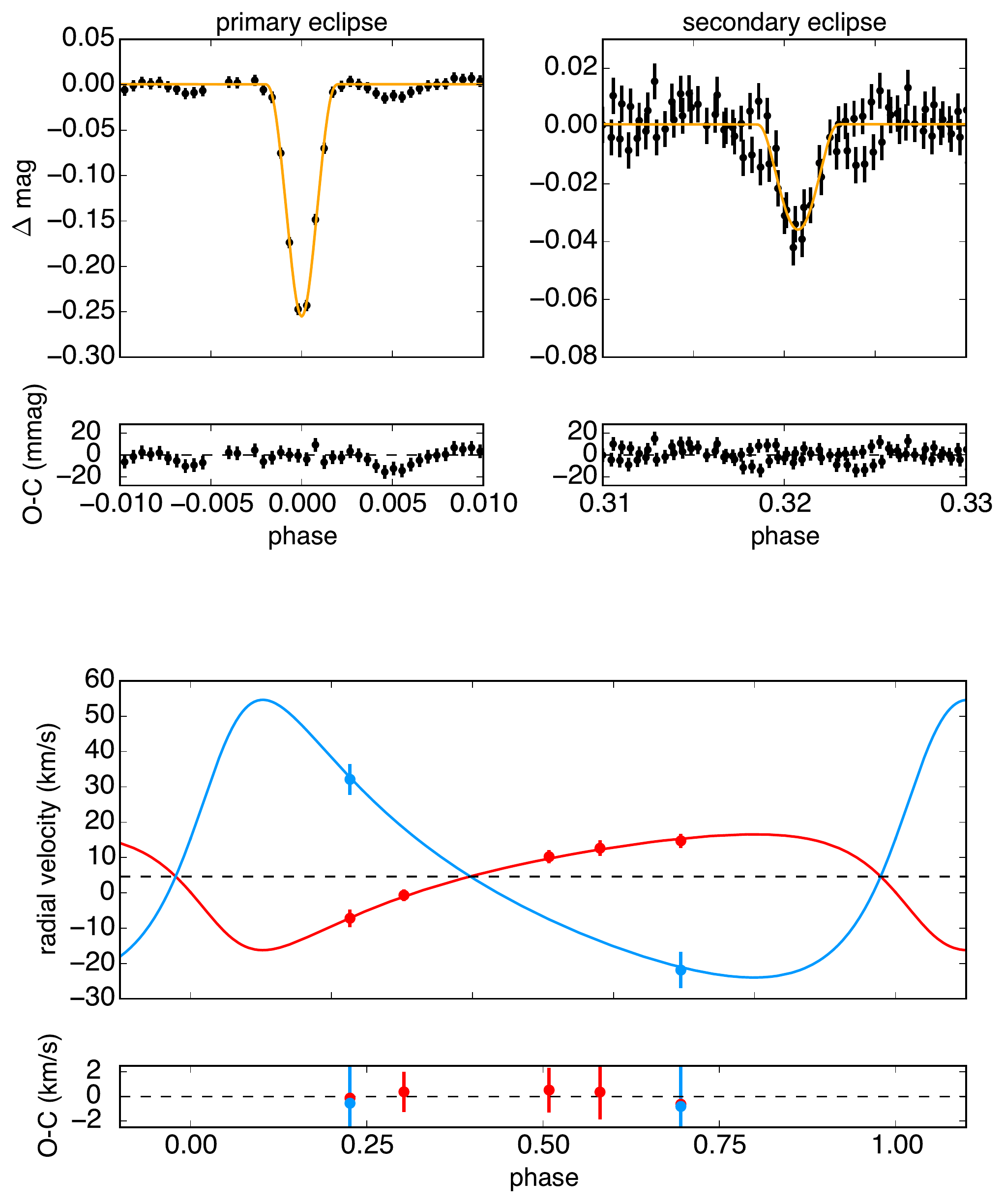}
\caption{JKTEBOP fit to the K2 light curve and RVs for MHO 9. The RV point near phase of 0.3 (corresponding to the UT 2016-01-24  observation) was not included in this analysis but is entirely consistent with our best fit.}
\label{fig:bpl116fit}
\end{figure}

The K2 light curve 
exhibits a single primary eclipse (at BJD 2457099 with a depth of $\sim 24\%$), and two shallow events of similar widths and shapes (at BJD 2457070 and 2457113 with depths of $\sim 5\%$).  Assuming that these two events are both secondary events of the same system, then we can determine the period to be $\sim 42.8$~d 
(see Fig.~\ref{fig:bpl116fit}).  
Note that all of the following analysis depends upon this assumption; further follow-up is necessary to confirm both the period of this system and the presence of these shallow secondary events.

From the phase separation between the primary and secondary eclipses, we can constrain the value for $e \cos \omega$ to be approximately $-0.29$.  The durations of the eclipses can be measured only approximately given the single primary event and poor signal-to-noise. However, doing so yields a constraint on $e \sin \omega$ and gives an estimate for eccentricity ($e \sim$ 0.29) and argument of periastron ($\omega \sim$ 3.2).

We used \jktebop\ to fit the K2 photometry and all RV measurements. For two epochs, the velocity separation between components was small enough that two peaks were not distinguishable in the CCF. In such cases, we take the derived velocity to be the RV of the primary and do not include any secondary velocities in the RV fitting procedure.  We assumed a linear limb darkening law with coefficient $u$=0.6 for each component, though because the eclipses are only partial and the ingress and egress are sparsely sampled, limb darkening can not be strongly constrained with current data. The resulting physical parameters are summarized in Table~\ref{table:bpl116table} and the best-fit model shown in Figure~\ref{fig:bpl116fit}.

We note that there is an apparent discrepancy between the positions of MHO 9 relative to HCG 76 in the color-magnitude diagram compared to the positions of these objects in the mass-radius diagram; while the nearly identical components of HCG 76 occupy a position of higher mass in the $(V-K_s)$-$V$ diagram (see Fig.~\ref{fig:CMD}), the primary of MHO 9 appears to be more massive than either of the components of HCG 76 in the mass-radius diagram (see Fig.~\ref{fig:pleiades-mr}). Furthermore, the components of MHO 9 appear younger (i.e. larger) in the mass-radius diagram.

We do not have a satisfactory explanation for this, though we do note that substantial uncertainties remain in the masses and radii for the MHO 9 system, owing to incomplete phase coverage in the RV curves, difficulties in extracting the secondary RVs due to small velocity separations relative to the spectrograph resolution, and only a single primary eclipse and two presumed secondary eclipses in the K2 photometry due to the long period of the system. Thus, within these large uncertainties the system is consistent with Pleiades age in the mass-radius diagram, and the primary is still possibly less massive than either of the components of HCG 76 at one end of its uncertainty range, and the secondary is possibly quite close to the hydrogen-burning limit (as shown in Fig.~\ref{fig:pleiades-mr}).  However, an alternate solution that forces the fit to produce a primary mass for MHO 9 that is lower than the primary mass for HCG 76 yields masses for MHO 9 of $M_1, M_2$ = 0.26, 0.12 \msun\ with $\sim$20-25\% uncertainty, and radii of $R_1=0.37, 0.29$ \rsun\ with $\sim$15\% uncertainty.

\renewcommand{\period}{42.80} 
\renewcommand{\eperiod}{(fixed)}
\renewcommand{\tzero}{2266.21943}
\renewcommand{\etzero}{0.00064}
\renewcommand{\sbratio}{1.04}
\renewcommand{\esbratio}{0.42}
\renewcommand{\sumradii}{0.0181}
\renewcommand{\esumradii}{0.0011}
\renewcommand{\ratradii}{0.70}
\renewcommand{\eratradii}{0.20}
\renewcommand{\inclination}{89.278}
\renewcommand{\einclination}{0.094}
\renewcommand{\ecosw}{-0.272}
\renewcommand{\eecosw}{0.019}
\renewcommand{\esinw}{0.302}
\renewcommand{\eesinw}{0.083}
\renewcommand{\rvprim}{16.4}
\renewcommand{\ervprim}{3.0}
\renewcommand{\rvsec}{39.3}
\renewcommand{\ervsec}{6.8}
\renewcommand{\rvsys}{4.6}
\renewcommand{\ervsys}{1.3}
\renewcommand{\rada}{0.0106}
\renewcommand{\erada}{0.0016}
\renewcommand{\radb}{0.00746}
\renewcommand{\eradb}{0.00095}
\renewcommand{\lightratio}{0.510}
\renewcommand{\elightratio}{0.097}
\renewcommand{\ecc}{0.406}
\renewcommand{\eecc}{0.056}
\renewcommand{\omegap}{132.0}
\renewcommand{\eomegap}{9.8}
\renewcommand{\bpri}{0.76}
\renewcommand{\ebpri}{0.26}
\renewcommand{\bsec}{1.41}
\renewcommand{\ebsec}{0.28}
\renewcommand{\rchisq}{1.038}
\renewcommand{\erchisq}{0.366}
\renewcommand{\rmslc}{6.320}
\renewcommand{\rmsprv}{0.451}
\renewcommand{\rmssrv}{0.693}
\renewcommand{\sma}{43.0} 
\renewcommand{\esma}{5.7}
\renewcommand{\mrat}{0.417} 
\renewcommand{\emrat}{0.075} 
\renewcommand{\mpri}{0.41} 
\renewcommand{\empri}{0.18} 
\renewcommand{\msec}{0.172} 
\renewcommand{\emsec}{0.069} 
\renewcommand{\rpri}{0.46} 
\renewcommand{\erpri}{0.11} 
\renewcommand{\rsec}{0.321} 
\renewcommand{\ersec}{0.060} 
\renewcommand{\logga}{4.73} 
\renewcommand{\elogga}{0.12} 
\renewcommand{\loggb}{4.66} 
\renewcommand{\eloggb}{0.14} 
\renewcommand{\rhoa}{4.3} 
\renewcommand{\erhoa}{2.4} 
\renewcommand{\rhob}{5.2} 
\renewcommand{\erhob}{1.6} 

\begin{deluxetable}{lrll}
\tabletypesize{\scriptsize}
\tablecaption{System Parameters of MHO 9 (BPL 116) \label{table:bpl116table}}
\tablewidth{0.99\textwidth}
\tablehead{
\colhead{Parameter} & \colhead{Symbol} & \colhead{\jktebop} & \colhead{Units}
\\
& & \colhead{Value} & 
} 

\startdata
Orbital period & $P$ & \period\  \eperiod & days \\
Ephemeris timebase - 2454833 & $T_0$ & \tzero\ $\pm$ \etzero & BJD \\
Surface brightness ratio & $J$ & \sbratio\ $\pm$ \esbratio &  \\
Sum of fractional radii & $(R_1+R_2)/a$ & \sumradii\ $\pm$ \esumradii & \\
Ratio of radii & $k$ & \ratradii\ $\pm$ \eratradii &  \\
Orbital inclination & $i$ & \inclination\ $\pm$ \einclination & deg \\
Combined eccentricity, periastron longitude & $e\cos\omega$ & \ecosw\ $\pm$ \eecosw & \\
Combined eccentricity, periastron longitude & $e\sin\omega$ & \esinw\ $\pm$ \eesinw & \\
Primary radial velocity amplitude & $K_1$ & \rvprim\ $\pm$ \ervprim & km s$^{-1}$ \\
Secondary radial velocity amplitude & $K_2$ & \rvsec\ $\pm$ \ervsec & km s$^{-1}$ \\
Systemic radial velocity & $\gamma$ & \rvsys\ $\pm$ \ervsys & km s$^{-1}$ \\
Fractional radius of primary & $R_1/a$ & \rada $\pm$ \erada & \\
Fractional radius of secondary & $R_2/a$ & \radb $\pm$ \eradb & \\
Luminosity ratio & $L_2/L_1$ & \lightratio\ $\pm$ \elightratio & \\
Eccentricity & $e$ & \ecc\ $\pm$ \eecc & \\
Periastron longitude & $\omega$ & \omegap\ $\pm$ \eomegap & deg \\
Impact parameter of primary eclipse & $b_1$ & \bpri\ $\pm$ \ebpri & \\
Impact parameter of secondary eclipse & $b_2$ & \bsec\ $\pm$ \ebsec & \\
Orbital semi-major axis & $a$ & \sma\ $\pm$ \esma & \rsun  \\
Mass ratio & $q$ & \mrat\ $\pm$ \emrat & \\
Primary mass & $M_1$ & \mpri\ $\pm$ \empri & \msun \\
Secondary mass & $M_2$ & \msec\ $\pm$ \emsec & \msun \\
Primary radius & $R_1$ & \rpri\ $\pm$ \erpri & \rsun \\
Secondary radius & $R_2$ & \rsec\ $\pm$ \ersec & \rsun \\
Primary surface gravity & $\log g_1$ & \logga\ $\pm$ \elogga & cgs \\
Secondary surface gravity & $\log g_2$ & \loggb\ $\pm$ \eloggb & cgs \\
Primary mean density & $\rho_1$ & \rhoa\ $\pm$ \erhoa & $\rho_\odot$ \\
Secondary mean density & $\rho_2$ & \rhob\ $\pm$ \erhob & $\rho_\odot$ \\
Temperature ratio & $T_2/T_1$ & 1.01 $\pm$ 0.10 & \\
Primary temperature & $T_1$ & 2970    &    K\\
Secondary temperature & $T_2$ & 3000   & K\\
Reduced chi-squared of light curve fit & $\chi^2_\mathrm{red}$ & \rchisq  & \\
RMS of best fit light curve residuals & & \rmslc  & mmag \\
RMS of primary RV residuals & & \rmsprv & km s$^{-1}$ \\
RMS of secondary RV residuals & & \rmssrv & km s$^{-1}$
\tablecomments{The \jktebop\ best-fit orbital parameters and 1-$\sigma$ uncertainties result from 1,000 Monte Carlo simulations. The primary temperature is calculated from the $V-K_s$ color, as described in \S~\ref{sec:stellarprops}.}
\enddata

\end{deluxetable}


\subsection{HD 23642}
\label{sec:hd23642}

HD 23642 (HII 1431, EPIC 211082420) is a known double-lined EB that has been well characterized in the literature, and in fact has been used to provide highly precise distances to the Pleiades \citep[e.g.][]{munari2004, southworth2005, groenewegen2007}. The system was first noted to be a double-lined spectroscopic binary by \cite{pearce1957} and was discovered to be eclipsing by \cite{torres2003} from Hipparcos epoch photometry. The primary spectral type from \cite{abt1978} is A0vp(Si)+Am. Though high quality ground-based photometric time series exist for this system, $K2$ has delivered the most extensive (covering 29 complete orbital phases) and precise light curve to date, despite being clearly saturated on the detector. With a period of $P\approx2.46$~d, the system clearly exhibits ellipsoidal modulation and reflection effects in the raw $K2$ photometry. 

We used \jktebop\ to mutually fit the new $K2$ photometry and literature radial velocities, providing direct determinations of the masses and radii. The photometry used for this purpose was the PDC SAP flux from the publicly available files on MAST. No additional treatment of the light curve was performed for this analysis. The 6.5-hr pseudo-CDPP\footnote{See \cite{aigrain2015} for a detailed definition of the pseudo-CDPP.} (combined differential photometric precision) across the entire light curve was taken as the constant observational error for each measurement. RVs were adopted from \cite{munari2004} and  \cite{groenewegen2007}. The \cite{groenewegen2007} RVs in their Table~2 are \emph{relative to systemic}, so to each measurement we added the final best-fit systemic RV from the PHOEBE fit presented in their Table~5. Additional RVs for this system exist in \cite{pearce1957} and \cite{abt1958}, but were excluded here due to their lower precision, following \cite{southworth2005}. 

In order to better constrain the ratio of radii, we imposed the following light ratio: $l_2/l_1$=0.354 $\pm$ 0.035. We calculated this light ratio for the $Kepler$ bandpass by convolving the $Kepler$ throughput curve\footnote{The $Kepler$ response function is available at \url{http://keplergo.arc.nasa.gov/kepler_response_hires1.txt}} with ATLAS9 model atmospheres of temperatures $T_1=9750$ K, $T_2=7600$ K \citep{southworth2005}, $\log{g}$=4.5 dex, $Z$=0, and scaling the flux ratio to equal the \cite{torres2003} value of $l_2/l_1$ = $0.31 \pm 0.03$ for a 45~\AA\ window centered on 5187~\AA. The uncertainty in our light ratio comes from assuming that the 10\% error measured by \cite{torres2003} is preserved when we calculated our synthetic value from model atmospheres.

We hold the eccentricity fixed at zero in our fit, consistent with prior studies of this system and with expectations of the tidal circularization timescale compared to the system age. The mass ratio used by \jktebop\ to calculate the out-of-eclipse variability due to ellipsoidal modulation was held fixed at the spectroscopic value obtained from an initial fit of the RVs, $q$=0.7030. As also noted by \cite{southworth2005}, inclusion of the reflection effect was required to obtain a good fit in the out-of-eclipse portions of the light curve. Initial estimates for the primary and secondary reflection coefficients were found by manually adjusting these parameters in the initial fitting stages until an acceptable fit was found. The reflection coefficients were then left as free parameters in the final fit. 
As prior authors have done, we fixed the gravity darkening exponent, $\beta$, to 1 for each component. This value is expected for such hot stars with radiative envelopes, but we note that our final solution favors oblateness values very close to zero (i.e., consistent with spherical stars) which is also supported by the modest $v\sin{i}$ values ($<40$ \kms) measured for both components.

A linear limb darkening law for both components was assumed, and we allowed the limb darkening coefficient, $u$, to be a free parameter for both stars, using appropriate values from \citet{claret2011} as initial estimates. Prior authors have not been able to constrain limb darkening in this system due to lower fidelity ground-based light curves. To account for the relatively long integration time of the photometry compared to the orbital period, we numerically integrated the models at 10 points in 1766 second intervals, corresponding to the length of $Kepler$ long cadence observations.

\begin{figure}[ht!]
\centering
\includegraphics[width=0.7\textwidth]{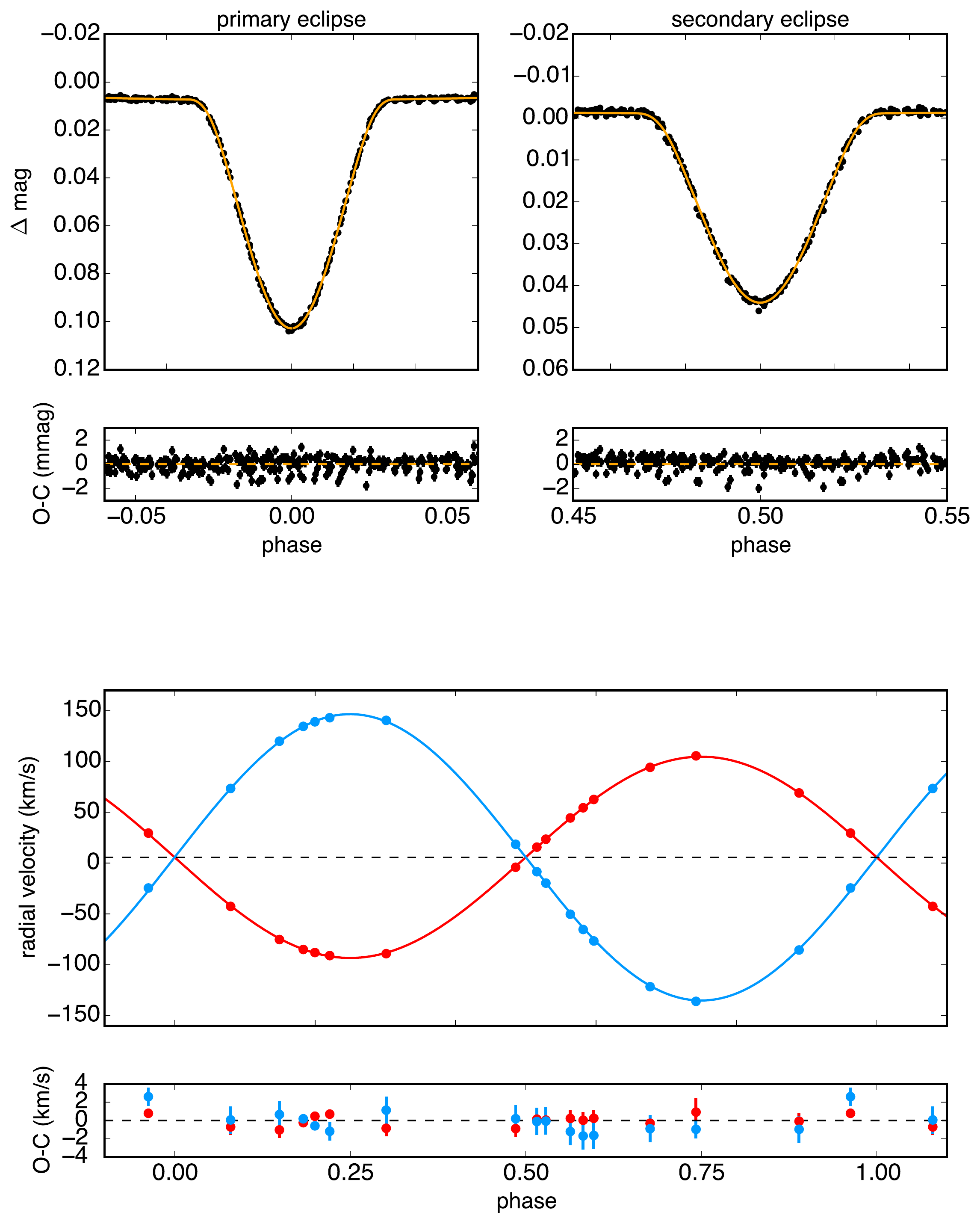}
\caption{Top panels: $K2$ PDC SAP light curve for HD 23642 phase folded on the orbital period of $\approx$2.46 days, with the best-fit \jktebop\ model plotted in orange. Bottom panel: Literature radial velocities from \cite{munari2004} and \cite{groenewegen2007} with the best-fit \jktebop\ models indicated by the red and blue curves. In each panel the best-fit residuals are plotted below.}
\label{fig:hd23642-fit}
\end{figure}

\begin{figure}[ht!]
\centering
\includegraphics[width=0.5\textwidth]{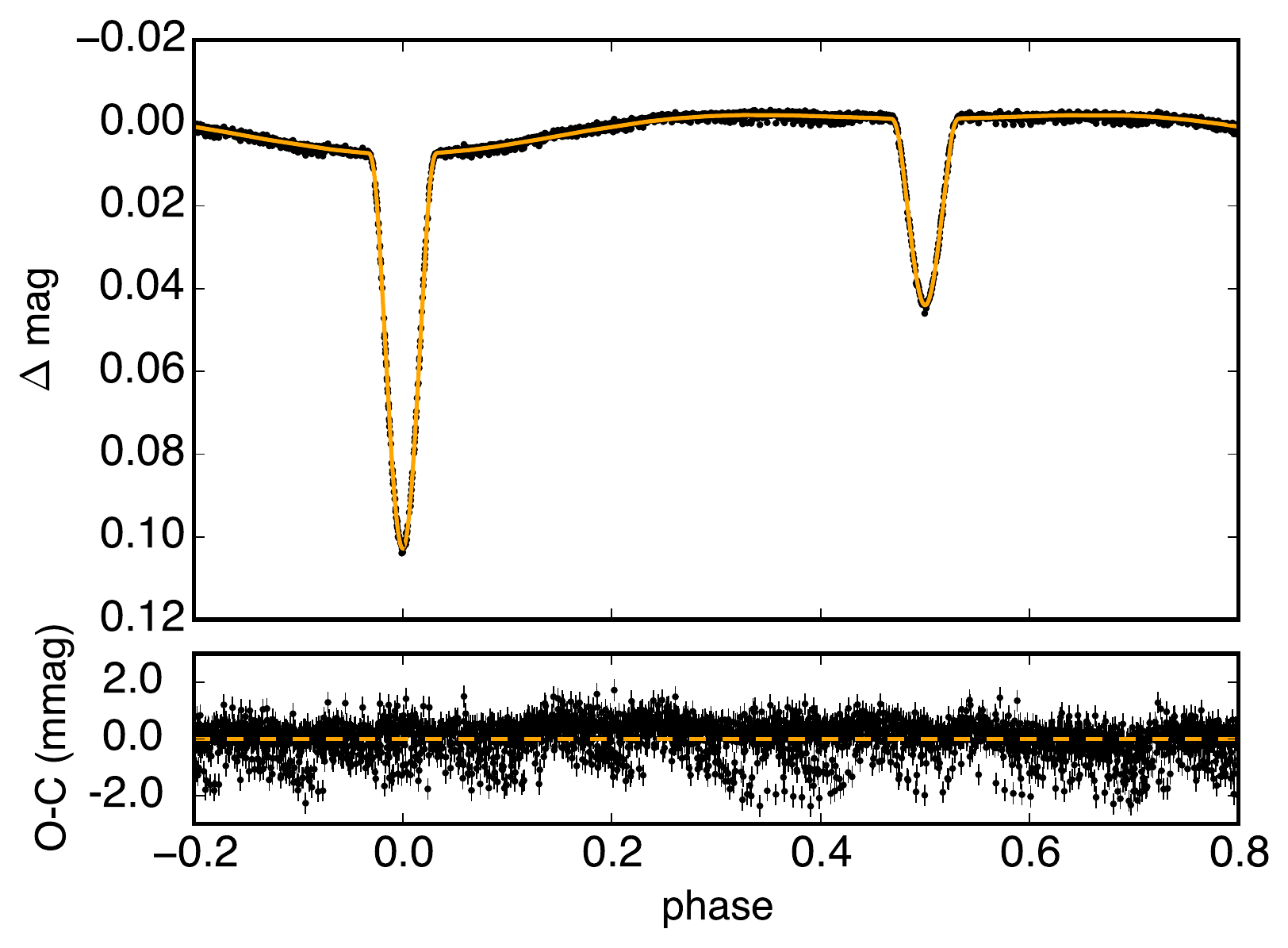}
\caption{The complete $K2$ light curve for HD 23642, phase folded on the best period and showing the best-fit \jktebop\ model in orange. The best fit residuals are plotted below.}
\label{fig:hd23642-fullfit}
\end{figure}

The best-fitting \jktebop\ model to the $K2$ photometry and literature RVs are depicted in Figures~\ref{fig:hd23642-fit}--\ref{fig:hd23642-fullfit} and the resulting parameters listed in Table~\ref{tab:hd23642table}. For comparison, in Table~\ref{tab:hd23642lit} we provide parameters from the literature for each study that has characterized this system in detail. We find masses for both components that are consistent with prior determinations in the literature. For the radii, we find a secondary radius that is consistent with the literature, but a primary radius that is $\sim$5\% smaller than previously reported values. Additionally, we find a temperature ratio that is higher than previous authors have found, though the absolute temperature of the secondary is consistent within error with prior determinations, assuming the primary temperature and uncertainty from \cite{southworth2005}. 

\begin{figure}[ht!]
\centering
\includegraphics[width=0.75\textwidth]{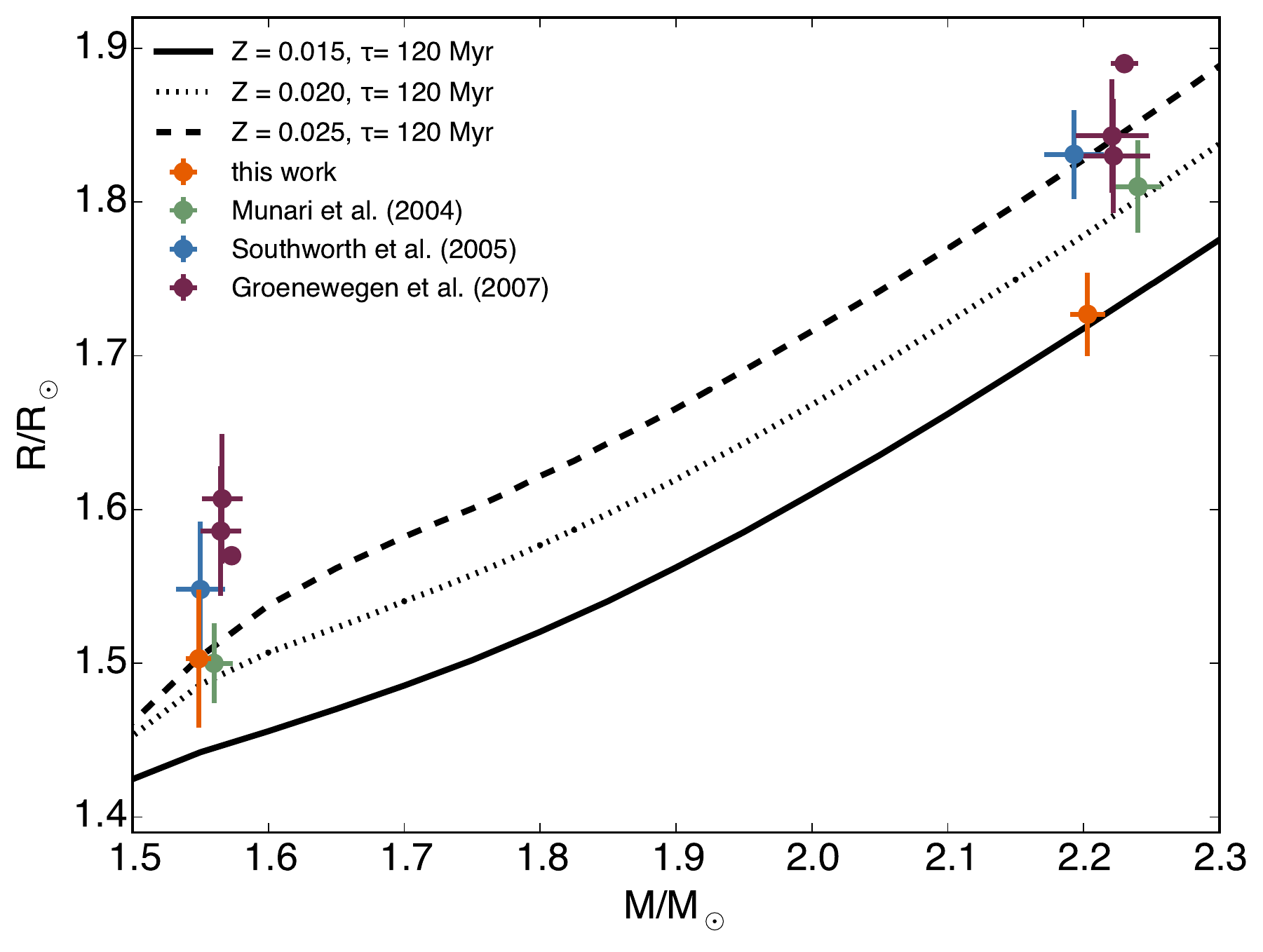}
\caption{PARSEC v1.2S 120 Myr isochrones \citep{bressan2012, chen2015} in the mass-radius plane, of different metallicities. Several direct determinations of the masses and radii of the double-lined EB HD 23642 from the literature are indicated, along with our new determinations from the highly precise K2 light curve and literature RVs.
}
\label{fig:hd23642-parsec}
\end{figure}

While it is surprising to find a radius for the primary star that is so discrepant from prior determinations, we show in Fig.~\ref{fig:hd23642-parsec} that in this case the primary is much closer to the Pleiades age isochrone in the mass-radius
plane when compared to the \cite{bressan2012} PARSEC v1.2S isochrones assuming the recently revised solar metallicity (Z$_\odot$=0.015) of \cite{caffau2011}. In fact, the primary parameters are entirely consistent with the accepted Pleiades age in both planes. All prior determinations of the primary parameters suggest the star is roughly a factor of two or more older than Pleiades age using this metallicity value. We also note that the fractional uncertainties in our mass and radius determinations are $\sim$2--4\%, which are consistent with \cite{munari2004, southworth2005} but inconsistent with \cite{groenewegen2007}. We suggest these last authors likely underestimated their mass and radius uncertainties. In all determinations, there is a significant degree of apparent non-coevality between the primary and secondary components, 
but is greatly ameliorated with our updated parameters and the degree of which is also somewhat lessened by adopting super-solar metallicities.

The positions of these stars in the mass-radius plane relative to evolutionary models has been studied previously in \cite{southworth2005}. Those authors invoked super-solar metallicities ($0.02 < Z < 0.03$) to reconcile the large radii with the Pleiades age. \cite{groenewegen2007} also made comparisons with evolutionary models, using a value of [Fe/H]=+0.058 dex. However, \cite{soderblom2009} measured the metallicity of the Pleiades from high-resolution echelle spectroscopy of 20 members, finding [Fe/H]= +0.03 $\pm$ 0.02 $\pm$ 0.05 dex (statistical and systematic errors quoted), which they compared to the average across previously published values of [Fe/H] = +0.042 $\pm$ 0.021 dex. Thus, the evolution models employed by prior authors to demonstrate the positions of these stars relative to Pleiades age isochrones may have represented the metal rich end of the true Pleiades metallicity distribution.

\renewcommand{\period}{2.46113408}
\renewcommand{\eperiod}{0.00000050}
\renewcommand{\tzero}{119.522070}
\renewcommand{\etzero}{0.000021}
\renewcommand{\sbratio}{0.4859}
\renewcommand{\esbratio}{0.0068}
\renewcommand{\sumradii}{0.2712}
\renewcommand{\esumradii}{0.0014}
\renewcommand{\ratradii}{0.870}
\renewcommand{\eratradii}{0.039}
\newcommand{\ldprim}{0.412}
\newcommand{\eldprim}{0.032}
\newcommand{\ldsec}{0.510}
\newcommand{\eldsec}{0.034}
\newcommand{\rflprim}{0.00246}
\newcommand{\erflprim}{0.00015}
\newcommand{\rflsec}{0.00648}
\newcommand{\erflsec}{0.00015}
\renewcommand{\inclination}{78.21}
\renewcommand{\einclination}{0.11}
\renewcommand{\rvprim}{99.02}
\renewcommand{\ervprim}{0.27}
\renewcommand{\rvsec}{140.86}
\renewcommand{\ervsec}{0.36}
\renewcommand{\rvsys}{5.68}
\renewcommand{\ervsys}{0.16}
\renewcommand{\rada}{0.1450}
\renewcommand{\erada}{0.0023}
\renewcommand{\radb}{0.1262}
\renewcommand{\eradb}{0.0037}
\renewcommand{\lightratio}{0.355}
\renewcommand{\elightratio}{0.035}
\renewcommand{\bpri}{1.409}
\renewcommand{\ebpri}{0.036}
\renewcommand{\bsec}{1.409}
\renewcommand{\ebsec}{0.036}
\renewcommand{\rchisq}{2.59}
\renewcommand{\rmslc}{0.61}
\renewcommand{\rmsprv}{0.59}
\renewcommand{\rmssrv}{1.12}
\renewcommand{\sma}{11.915} 
\renewcommand{\esma}{0.023} 
\renewcommand{\mrat}{0.7030} 
\renewcommand{\emrat}{0.0027} 
\renewcommand{\mpri}{2.203} 
\renewcommand{\empri}{0.013} 
\renewcommand{\msec}{1.5488} 
\renewcommand{\emsec}{0.0093} 
\renewcommand{\rpri}{1.727} 
\renewcommand{\erpri}{0.027} 
\renewcommand{\rsec}{1.503} 
\renewcommand{\ersec}{0.045} 
\renewcommand{\logga}{4.306} 
\renewcommand{\elogga}{0.014} 
\renewcommand{\loggb}{4.274} 
\renewcommand{\eloggb}{0.025} 
\renewcommand{\rhoa}{0.427} 
\renewcommand{\erhoa}{0.020} 
\renewcommand{\rhob}{0.456} 
\renewcommand{\erhob}{0.040} 

\begin{deluxetable}{lrll}
\tabletypesize{\scriptsize}
\tablecaption{ System Parameters of HD 23642 \label{tab:hd23642table}}
\tablewidth{0.99\textwidth}
\tablehead{
\colhead{Parameter} & \colhead{Symbol} & \colhead{\jktebop} & \colhead{Units}
\\
& & \colhead{Value} &
} 

\startdata
Orbital period & $P$ & \period\ $\pm$ \eperiod & days \\
Ephemeris timebase - 2457000 & $T_0$ & \tzero\ $\pm$ \etzero & BJD \\
Surface brightness ratio & $J$ & \sbratio\ $\pm$ \esbratio &  \\
Sum of fractional radii & $(R_1+R_2)/a$ & \sumradii\ $\pm$ \esumradii & \\
Ratio of radii & $k$ & \ratradii\ $\pm$ \eratradii  & \\
Orbital inclination & $i$ & \inclination\ $\pm$ \einclination & deg \\
Primary limb darkening coefficient & $u_1$ & \ldprim $\pm$ \eldprim &  \\
Secondary limb darkening coefficient & $u_2$ & \ldsec $\pm$ \eldsec &  \\
Primary geometric reflection coefficient & $r_1$ & \rflprim $\pm$ \erflprim &  \\
Secondary geometric reflection coefficient & $r_2$ & \rflsec $\pm$ \erflsec &  \\
Primary radial velocity amplitude & $K_1$ & \rvprim\ $\pm$ \ervprim & km s$^{-1}$ \\
Secondary radial velocity amplitude & $K_2$ & \rvsec\ $\pm$ \ervsec & km s$^{-1}$ \\
Systemic radial velocity & $\gamma$ & \rvsys\ $\pm$ \ervsys & km s$^{-1}$ \\
Fractional radius of primary & $R_1/a$ & \rada $\pm$ \erada & \\
Fractional radius of secondary & $R_2/a$ & \radb $\pm$ \eradb & \\
Luminosity ratio & $L_2/L_1$ & \lightratio\ $\pm$ \elightratio & \\
Impact parameter of primary eclipse & $b_1$ & \bpri\ $\pm$ \ebpri & \\
Impact parameter of secondary eclipse & $b_2$ & \bsec\ $\pm$ \ebsec & \\
Orbital semi-major axis & $a$ & \sma\ $\pm$ \esma & \rsun  \\
Mass ratio & $q$ & \mrat\ $\pm$ \emrat & \\
Primary mass & $M_1$ & \mpri\ $\pm$ \empri & \msun \\
Secondary mass & $M_2$ & \msec\ $\pm$ \emsec & \msun \\
Primary radius & $R_1$ & \rpri\ $\pm$ \erpri & \rsun \\
Secondary radius & $R_2$ & \rsec\ $\pm$ \ersec & \rsun \\
Primary surface gravity & $\log g_1$ & \logga\ $\pm$ \elogga & cgs \\
Secondary surface gravity & $\log g_2$ & \loggb\ $\pm$ \eloggb & cgs \\
Primary mean density & $\rho_1$ & \rhoa\ $\pm$ \erhoa & $\rho_\odot$ \\
Secondary mean density & $\rho_2$ & \rhob\ $\pm$ \erhob & $\rho_\odot$ \\
Reduced chi-squared of light curve fit & $\chi^2_\mathrm{red}$ & \rchisq  & \\
RMS of best fit light curve residuals & & \rmslc &  mmag \\
Reduced chi-squared of primary RV fit & $\chi^2_\mathrm{red}$ & 0.63  &  \\
RMS of primary RV residuals & & \rmsprv & km s$^{-1}$ \\
Reduced chi-squared of secondary RV fit & $\chi^2_\mathrm{red}$ & 0.95  & \\
RMS of secondary RV residuals & & \rmssrv & km s$^{-1}$
\tablecomments{Best-fit orbital parameters and their uncertainties resulting from 1,000 Monte Carlo simulations with \jktebop. For this fit the eccentricity was fixed at zero, and the gravity darkening exponents for both the primary and secondary were fixed at one.}
\enddata

\end{deluxetable}

\begin{deluxetable}{ccccc} 
\tabletypesize{\footnotesize} 
\tablewidth{0.99\textwidth} 
\tablecaption{ Parameters derived for HD 23642 from the literature \label{tab:hd23642lit}} 
\tablehead{ 
\colhead{Parameter} & \colhead{Groenewegen et al. (2007)} & \colhead{Southworth et al. (2005)} & \colhead{Munari et al. (2004)} & \colhead{Torres (2003)}
}
\startdata 
$P$ (d) &  2.46113358 $\pm$ 0.00000015 & fixed at M04 value &  2.46113400 $\pm$ 0.00000034 &  2.46113329 $\pm$ 0.00000066 \\
$T_0$ (HJD) & 2452903.60002 $\pm$ 0.00014 & fixed at M04 value &  2452903.5981 $\pm$ 0.0013 & 2436096.5204 $\pm$ 0.0040 \\
$\gamma$ (\kms) & 5.39 $\pm$ 0.04 & 6.07 $\pm$ 0.39 & 5.17 $\pm$ 0.24  &  6.1 $\pm$ 1.7 \\
$q$ &  0.7054 $\pm$ 0.0006  & 0.7068 $\pm$ 0.0050 & 0.6966 $\pm$ 0.0034 &  0.6934 $\pm$ 0.0077 \\
$i$ (deg) & 76.63 $\pm$ 0.02 & 77.78 $\pm$ 0.17 & 78.10 $\pm$ 0.21 & $\sim$78 \\
$a$ (\rsun) &  11.959 $\pm$ 0.0052 & 11.906 $\pm$ 0.041 & 11.956 $\pm$ 0.030 & $\sim$11.82 \\
$e$ & 0.0 (fixed) & 0.0 (fixed) & 0.0 $\pm$ 0.002 &  0 (fixed) \\
$T_1$ (K) & 9950 (fixed) & 9750 $\pm$ 250 & 9671 (fixed) & \\
$T_2$ (K) & 7281 $\pm$ 9 &  7600 $\pm$ 400 & 7500 $\pm$ 61 & \\
$R_1$ (\rsun) & 1.890 $\pm$ 0.003 & 1.831 $\pm$ 0.029 &  1.81 $\pm$ 0.030 & \\
$R_2$ (\rsun) & 1.570 $\pm$ 0.003 & 1.548 $\pm$ 0.044 &  1.50 $\pm$ 0.026  & \\
$M_1$ (\msun) & 2.230 $\pm$ 0.010 & 2.193 $\pm$ 0.022 &  2.24 $\pm$ 0.017 & \\
$M_2$ (\msun) & 1.573 $\pm$ 0.002 & 1.550 $\pm$ 0.018 & 1.56 $\pm$ 0.014 & \\
$\log{g_1}$ (cgs) & 4.2331 $\pm$ 0.0024 & 4.254 $\pm$ 0.014 & 4.27 $\pm$ 0.015 & \\
$\log{g_2}$ (cgs) & 4.2426 $\pm$ 0.0018\tablefootnote{We calculated the surface gravity of the primary and secondary based on the masses and radii of the final fit from \cite{groenewegen2007}, as those authors did not report the values. The uncertainties come from Monte Carlo error propagation of the associated uncertainties in mass and radius.} & 4.249 $\pm$ 0.025 & 4.28 $\pm$ 0.016 & 
\enddata 
\end{deluxetable}


\subsection{HII 2407}

HII 2407 (EPIC 211093684) is a recently recognized EB in the Pleiades, consisting of a K2 type primary and likely M-type secondary.  The system was found to be a single-lined spectroscopic binary by \citet{mermilliod1992}. It was discovered as eclipsing in the K2 data and has been recently analyzed and discussed in \citet{david2015a}.  The K2 light curve is suggestive of a radius ratio of 0.27 and a large $\Delta T_\mathrm{eff}$ between the primary and secondary, consistent with the non-detection of secondary lines in optical spectra to date. Follow-up spectroscopy in the infrared, where the flux ratio between the primary and secondary is more favorable, is underway (L.~Prato, private communication) and should allow for the unique determination of the masses and radii of this system.


\subsection{AK II 465}

The star AK II 465 (EPIC 210822691) is an EB with possible membership to the Pleiades. The star was first mentioned, and given the AK classification, in the \cite{ak1970} proper motion survey of the Pleiades. \cite{mermilliod1997} classified the source as a non-member based on two RV measurements separated by 1065 days, yielding a mean RV of 
23.4 $\pm$ 0.4 \kms, compared to the mean Pleiades systemic radial velocity of $\sim$5 \kms. \cite{mermilliod2009} again asserted non-membership based on the same data, and additionally provided a $v\sin{i}$ measurement of 3.9 $\pm$ 1.9 \kms. However, given the binary nature of this system and short orbital period implied by the K2 light curve, two epochs may not be enough to invalidate membership on the basis of mean RV alone. The source is not discussed elsewhere in the literature.

Proper motion measurements for this source are inconsistent with Pleiades membership, as detailed in Table~\ref{table:akii465pm}. For reference, the Pleiades mean proper motion is $\mu_{\alpha}$, $\mu_{\delta}$ = 20.10, -45.39 mas yr$^{-1}$ \citep{vanleeuwen2009}. In particular, \cite{bouy2015} assigned membership probabilities to the source of $\lesssim$1\% based on the proper motions measured from either DANCe or Tycho-2 data sets.
In addition, in a $V$ versus $V-K$ CMD the source falls slightly below the Pleiades single star main sequence. The primary and secondary eclipse depths are similar, suggesting a temperature ratio (and thus mass ratio) close to unity. Thus, one would expect this system to be overluminous for its position in a CMD. Furthermore, the out-of-eclipse K2 light curve for the source is much less variable than any of the well-known Pleiades members of a similar $V$ magnitude. The evidence above is suggestive that AK II 465 is most likely a non-member. 

\begin{deluxetable}{llll}
\tablecolumns{4}
\tablecaption{Literature proper motion measurements for AK II 465\label{table:akii465pm}}
\tablehead{
    \colhead{$\mu_\alpha$ (mas yr$^{-1}$)} & \colhead{$\mu_\delta$ (mas yr$^{-1}$)} & 
    \colhead{Note} & 
    \colhead{Source}
}
\startdata
9.0$\pm$5.5 & -28.6$\pm$5.5 & URAT1 & \cite{zacharias2015}
\\
3.59$\pm$6.07 &	-22.05$\pm$6.07 & DANCe & \cite{bouy2015}
\\
4.8$\pm$0.7 & -37.5$\pm$0.7 & UCAC4 & \cite{zacharias2013}
\\
7.0$\pm$1.3 & -38.5$\pm$1.4 & PPMXL & \cite{roeser2010}
\\
8.3$\pm$1.4 & -39.3$\pm$1.5 & Tycho-2 & \cite{hog2000}
\enddata
\end{deluxetable}

Keck/HIRES spectra revealed the source to be double-lined, and from these spectra we estimate a G0 spectral type. Both components possess lithium absorption (see Table~\ref{table:akii465ew}), normally an indicator of extreme stellar youth; however, lithium may be removed from the stellar photosphere either via convective transport to depths hot enough for burning (the dominant mechanism for late G and K dwarfs) or due to gravitational settling (which is thought to be the case for F dwarfs). Around a spectral type of G0, neither process works significantly, and one expects lithium abundances that nearly reflect the local interstellar medium values for a wide range of ages \citep[see e.g.][]{soderblom1999, jones1999}. Thus, at this temperature or mass, lithium is not a useful youth diagnostic. Furthermore, \cite{barrado1996} showed that tidally locked binaries (as AK II 465 is expected to be) tend to retain lithium longer than both single stars or more widely separated binaries of the same age. 

\begin{deluxetable}{lll}
\tablecolumns{3}
\tablecaption{Keck/HIRES equivalent widths for AK II 465 \label{table:akii465ew}}
\tablehead{ 
    \colhead{Component} &
    \colhead{EW(Li I 6707.8) [\AA]} & \colhead{EW(Ca I 6717) [\AA]}
}
\startdata
A & 0.05 & 0.04 \\ 
B & 0.03 & 0.02
\enddata
\end{deluxetable}

From simultaneous fitting of the K2 photometry and four epochs of double-lined RV measurements, we confirmed that the systemic velocity of AK II 465 ($\gamma \sim$ 20 \kms) is inconsistent with the mean Pleiades motion. Furthermore, the derived masses and radii are consistent with an age of $\sim$4-5 Gyr when compared to BHAC15 models. Thus, we conclude that this EB is unlikely to be a true member of the Pleiades and indeed likely not a system of Pleiades age despite the appearance of modest Li in the spectrum. We provide the best fit parameters in Appendix~\ref{sec:akiiappendix} for completeness but do not discuss this EB further in the context of the Pleiades or Hyades below.


\subsection{vA 50 (HAN 87)}

vA 50 (HAN 87, EPIC 210490365) was first identified as a proper motion member of the Hyades by \citet{vanaltena1966}, under the name vA 50, using photographic plates from the Lick Observatory 20" astrographic telescope.   A subsequent proper motion survey, also using Lick plates, reconfirmed it as a candidate Hyades member based on its proper motion, with the additional name of HAN 87 \citep{hanson1975}. Concurrent with our identification and follow-up of the system, \citet{mann2016} recently reported vA 50 to harbor a Neptune-sized planet.

We estimated the physical parameters of the primary star as described in Section~\ref{sec:stellarprops}. \citet{reid1993} lists vA 50 as having an absolute $V$ magnitude of 15.70 and $V-I_C$ color of 2.91.  \citet{upgren1985} lists an absolute $V$ magnitude of 15.80 and a $V-I_K$ color of 2.87 which converts to $V-I_C$ of 2.95.  Adopting $V=15.80$ and $V-I_C=2.93$ gives $V-K=5.36$ and $T_\mathrm{eff} = 3170$~K and a spectral type of $\sim$M4. 

From a measured secular parallax of $22.3$ mas \citep{roser2011} based on proper motions from PPMXL \citep{roeser2010}, the distance can be estimated to be $\approx$44.8~pc.  We also find an estimated mass of $0.261 M_\odot$ and estimated radius of $0.321 R_\odot$ (see Section \ref{sec:stellarprops}). These values may be compared with those determined in the analysis by \citet{mann2016}, which estimates the host star to have a mass of $M_* = 0.294 \pm 0.021 M_\odot$, a radius of $R_* = 0.295 \pm 0.020 R_\odot$, and $T_\mathrm{eff} = 3180 \pm 60$~K.

From the BLS periodogram of the rectified light curve, we find that vA 50 exhibits a triangular eclipse shape every $\sim 3.48$~d with a width of approximately $0.04$~d and a depth of $1\%$.  It is not clear from the light curve alone if these events are transits, primary eclipses alone, or primary and secondary eclipses.  If this system is a stellar binary, then the period could be either $\sim 3.48$~d or $\sim 6.97$~d.  RV measurements or confirmation that the companion is indeed sub-stellar is necessary to distinguish between these two possibilities.

\begin{figure}[ht!]
\centering
\includegraphics[width=0.75\textwidth]{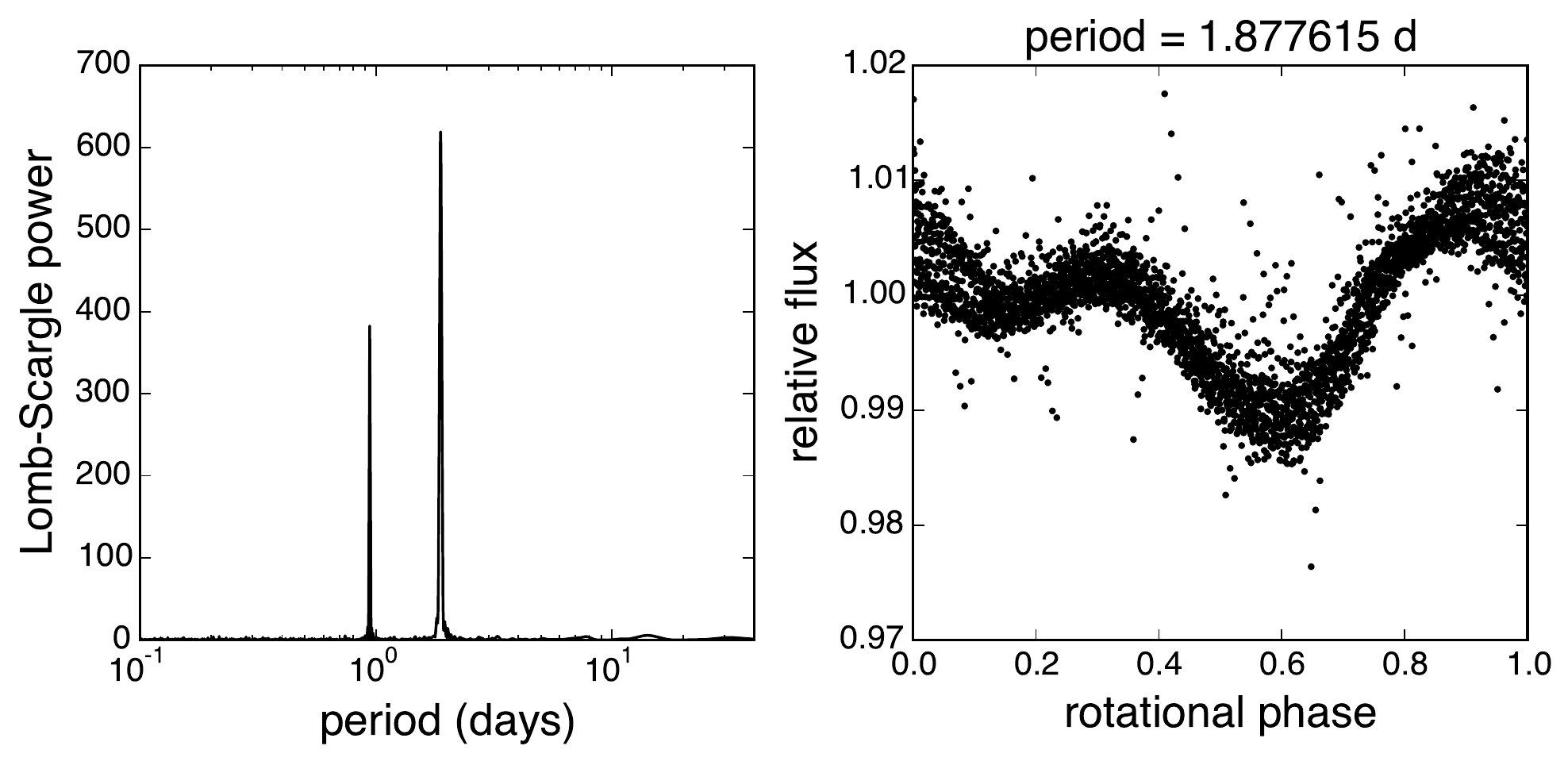}
\caption{Lomb-Scargle periodogram (left) and K2 PDC light curve phase folded on the favored rotation period of vA 50 (right). The other significant peak in the periodogram is at the half-period alias.}
\label{fig:h87-rotation}
\end{figure}

From the PDC light curve and a Lomb-Scargle periodogram analysis, we also measured the rotation period of the primary star from the repeating spot pattern (Figure~\ref{fig:h87-rotation}). We find $P_\mathrm{rot}$=1.88 $\pm$ 0.05~d, where the uncertainty has been approximated from the FWHM of the periodogram peak. Our rotation period is in agreement with the value reported in \cite{mann2016}. Those authors also reported a $v\sin{i}$ measurement of 7.8$\pm$0.5 \kms. Assuming an edge-on inclination, the rotational velocity and period imply a stellar radius of $R_* \approx 0.29 \pm 0.02$ \rsun, which is consistent with the \citet{mann2016} value, but slightly smaller than the value based on photometry quoted above.

\begin{figure}[ht!]
\centering
\includegraphics[width=0.75\linewidth]{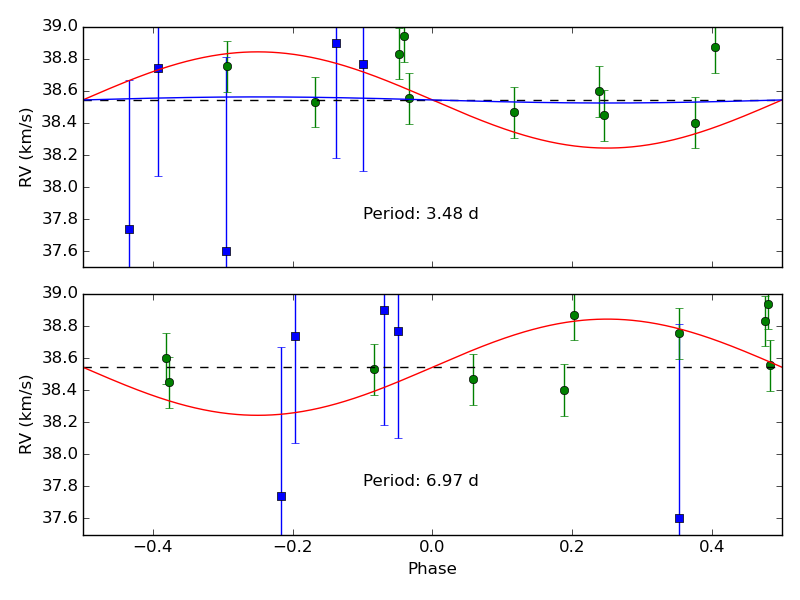}
\caption{Measured RVs of vA 50 (HAN 87, EPIC 210490365) folded on the two possible periods of $\sim 3.48$ and $\sim 6.97$ days, respectively.  RVs from \citet{mann2016} are plotted with green circles while our reported RVs are plotted with blue squares.  The dashed horizontal lines represent the median velocity of all RVs and is therefore assumed as the systemic velocity.  The red curve represents a circular orbit with an amplitude of 0.3 km/s.  This is not a fit, but rather a representative of the approximate maximum amplitude allowed by an RV curve to still be consistent with observations.  This limit is used in Figure~\ref{fig:h87_m2} to constrain the maximum mass of the companion for each of these periods and ultimately rule out a stellar companion.  The blue curve on the left panel represents the maximum amplitude of the estimated mass of the planetary companion.}
\label{fig:h87_rvs}
\end{figure}

\citet{mann2016} report 10 single-lined RV measurements using the IGRINS infrared spectrometer from the 2.7m Harlan J.\ Smith telescope at McDonald Observatory.  These measurements in addition to our 5 Keck single-lined RV measurements (Table~\ref{table:rvs}) result in full phase-coverage at either potential period that restricts the RV amplitude to be less than $0.3$ \kms\ (Figure~\ref{fig:h87_rvs}).  
If we take the color-estimated mass of $0.261 M_\odot$ to be the mass of the primary component of the system, then we can use this estimated maximum RV amplitude to constrain the mass of the companion.  Kepler's Third Law with either of the two possible periods constrains the semi-major axis of the system to be dependent only on the mass of the companion.  If we then assume a circular orbit and the most conservative case of an edge-on system ($i = 90^\circ$), then we can investigate the dependence of the RV semi-amplitude on the mass of the companion.  

\begin{figure}[ht!]
\centering
\includegraphics[width=0.75\linewidth]{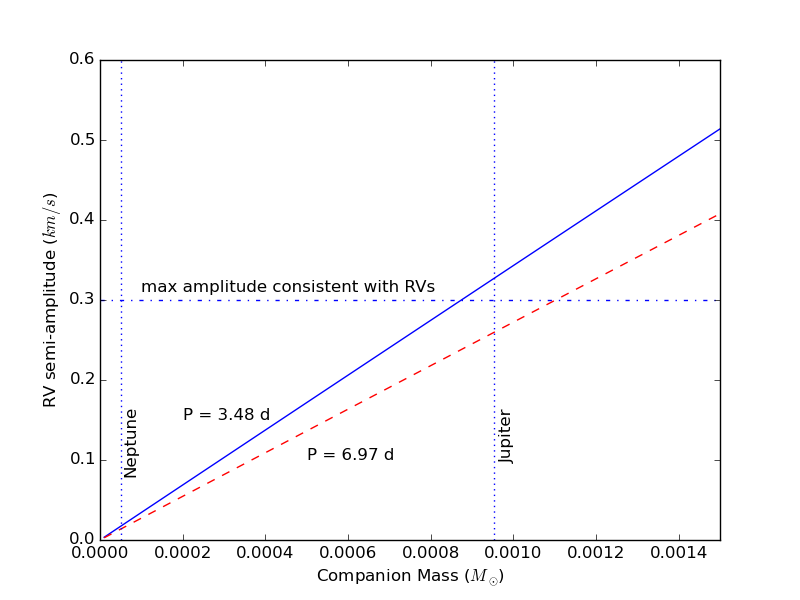}
\caption{Expected RV semi-amplitude as a function of the mass of the companion, assuming $M_1 = 0.261 M_\odot$ and $i=90^\circ$, for both possible periods (3.48 days shown as a solid blue line, 6.97 days shown as a dashed red line).  Dotted vertical lines represent the masses of Neptune and Jupiter, respectively.  The dot-dashed horizontal line depicts the approximate maximum amplitude that would be consistent with the measured RVs shown in Figure~\ref{fig:h87_rvs}.  This clearly rules out a stellar-companion, which also implies that the true period is in fact 3.48~d.}
\label{fig:h87_m2}
\end{figure}

Figure~\ref{fig:h87_m2} shows this relation along with the maximum amplitude consistent with the existing RV observations.  For either period, the mass of the companion must not be over $0.0011 M_\odot$ ($1.15 M_\mathrm{jup}$). 
Furthermore, since this would imply a sub-stellar companion, then the existence of a secondary eclipse is unlikely, strengthening the claim that we are seeing transits, rather than eclipses, at a period of $\sim 3.48$~d.

\begin{figure}[ht!]
\centering
\includegraphics[width=0.75\textwidth]{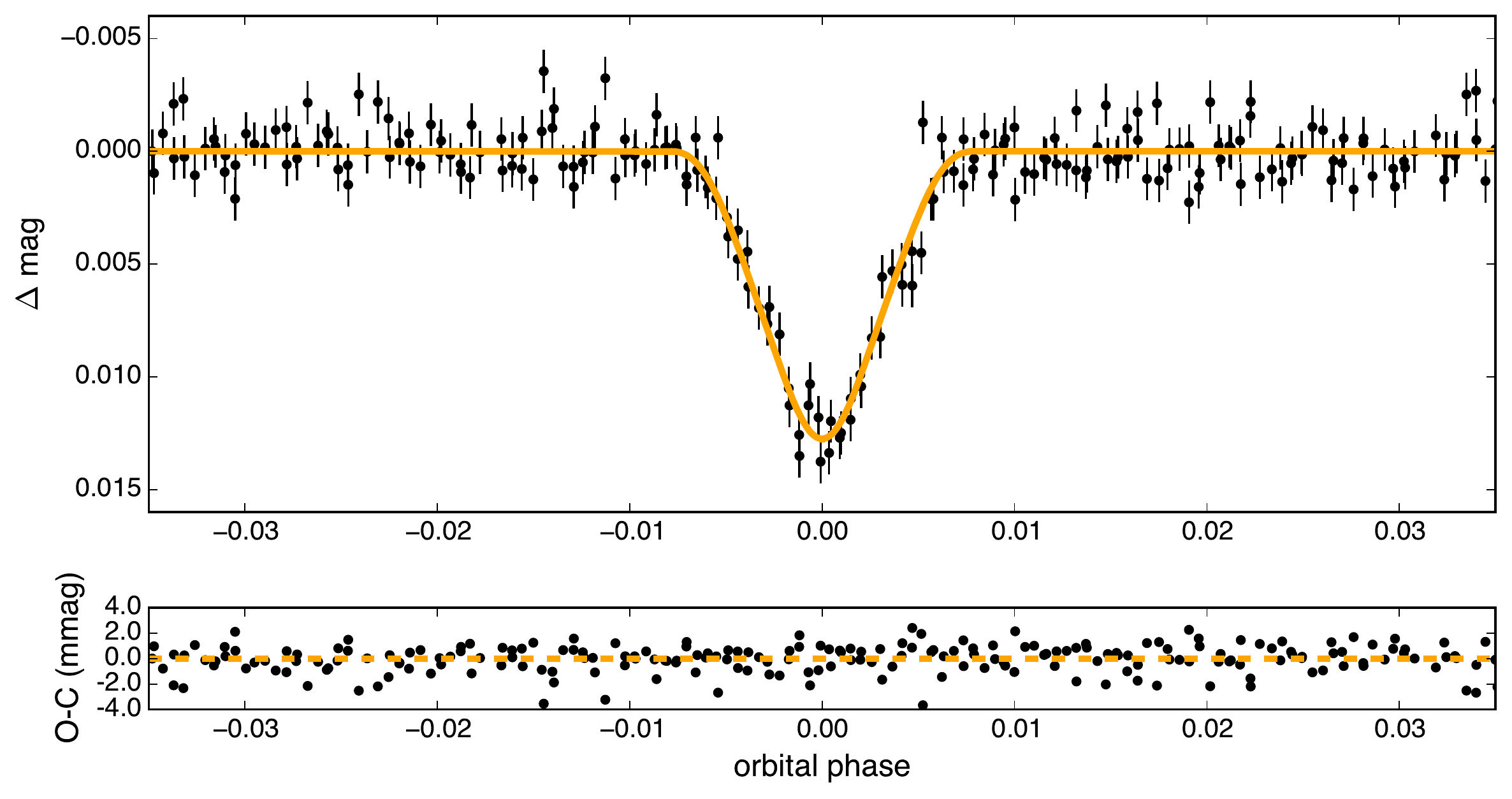}
\caption{Phase folded K2 light curve of vA 50 with the best-fitting \jktebop\ transit model shown in orange. The fit residuals are shown in the bottom panel.}
\label{fig:h87-ebop}
\end{figure}

Under the assumption of a sub-stellar object transiting, a \jktebop\ model was fit to the light curve data alone. Though originally designed to model EB light curves, \jktebop\ has also been demonstrated to reliably model exoplanet transits \citep[see e.g.][and references therein]{southworth2012}. The results and error estimates are shown in Table~\ref{table:h87table} with the best-fit model shown in Fig.~\ref{fig:h87-ebop}.  Most notably, the ratio of radii is estimated to be $R_p/R_* = 0.111$.  This along with the estimated radius of the host star of $R_* = 0.32 R_\odot$, gives the planetary companion a radius of $R_p = 0.035 R_\odot = 0.354 R_\mathrm{jup} = 1.01 R_\mathrm{nep}$.  
Under the assumption that this planetary companion has a density comparable to that of Neptune ($0.287 M_\odot/R_\odot^3$), its mass would be approximately $1.05 M_\mathrm{nep}$.  This mass would give a RV semi-amplitude of $< 0.019$ \kms\ which is consistent with the measured RV observations (Figs.~\ref{fig:h87_rvs}--\ref{fig:h87_m2}).
\citet{mann2016} also explore several possibilities besides a transiting Neptune, and also conclude that all scenarios involving an EB (blend, grazing, or companion EB) are inconsistent with the data. We note that the MCMC fitting results do admit solutions with companion radii as large as $\sim$0.2 \rsun\ or $\sim$2 \rjup. However, at the nominal age of the Hyades such a large radius would put the companion in the stellar mass regime, which is ruled out by the RVs.

We used the PDC light curve, subject to additional custom detrending via the procedure outlined in \citet{david2016}, for the purposes of fitting the K2 transits of vA 50b. We performed two fits, using \jktebop\ to model the transit curves in both cases. In the first fit, we used the \jktebop\ Levenberg-Marquardt fitting routine to find a best fit, determining parameter uncertainties through 1000 Monte Carlo simulations, as decribed in \S~\ref{subsec:modeling}. For the second fit, we employed the exact approach described in \citet{crossfield2015}, which uses standard minimization routines and the \texttt{emcee} Markov Chain Monte-Carlo (MCMC) implementation in Python \citep{foreman-mackey2013} to generate and assess the likelihoods of \jktebop\ model light curves, more fully exploring the degenerate parameter space \citep[see][for further details regarding burn-in treatment, chain initialization, and convergence testing]{crossfield2015}. For the MCMC fit, we assumed a Gaussian prior on the linear limb darkening parameter $u$, with $\mu=0.6$, $\sigma=0.1$, encompassing any reasonable value predicted by \citet{claret2012} for a star with temperature and surface gravity similar to vA 50. In the MCMC fit, we also allowed for modest eccentricity by imposing a Gaussian prior on $e$ with $\mu$=0.0, $\sigma$=0.01. \citet{mann2016} explored a solution with eccentricity as a free parameter, but in general the resulting parameters were consistent within error with the circular solution, and moreover the eccentricity is poorly constrained given the RV precision is not high enough to detect orbital motion at this stage. Thus, we report only a circular solution in our transit modeling analysis.

Our upper limit for the companion mass is consistent with both observational evidence and theoretical considerations that suggest Jovian mass planets should be rare around M-dwarfs; RV surveys have found that giant planets ($m\sin{i} \sim$ 0.3--3 \mjup) with orbital periods between 1--10~d are extremely scarce around M-dwarfs \citep{bonfils2013}, and core accretion is believed to be ineffective at forming such massive planets around low-mass stars, though Neptunes and super-Earths are thought to be more common \citep{laughlin2004}. For comparison, the MEarth project measured the occurrence rate of warm Neptunes transiting mid-to-late M dwarfs to be $<$0.15 per star \citep{berta2013}.

From the orbital period and the stellar mass adopted above, using Kepler's third law we estimate the separation of this putative planet to be $a \sim 0.03$ AU, well within the predicted location of the snow line ($\sim$1 AU) for a low-mass star at the time of gas disk dispersal \citep{kennedy2007}. If validated, this system would make an excellent target for future transit transmission spectroscopy studies, which would allow for a direct measurement of the C/O ratio in the planetary atmosphere, indicating where in the protoplanetary disk the planet may have formed \citep{oberg2011}.

\begin{deluxetable}{ccc}
\tabletypesize{\footnotesize}
\tablecaption{Fit to the K2 transits of vA 50b \label{table:h87table}}
\tablehead{ 
\colhead{Parameter} & 
\colhead{JKTEBOP value} & 
\colhead{MCMC fit value}
} 
\startdata

$P$ (days) & 3.48451 $\pm$ 0.00004 & $3.484505^{+0.000049}_{-0.000049}$ \\
$T_0$ (BJD-2457000) & 62.5801 $\pm$ 0.0005 & $62.58012^{+0.00056}_{-0.00053}$ \\
($R_P+R_*$)/$a$ & $0.0474^{+0.0099}_{-0.0082}$ & $0.047^{+0.039}_{-0.006}$ \\
$R_P/R_*$ & $0.111^{+0.020}_{-0.016}$ & $0.120^{+0.598}_{-0.006}$ \\
$i$ ($^\circ$) & $88.10^{+0.63}_{-0.44}$ & $88.0^{+0.4}_{-2.5}$ \\
$e\cos\omega$ & 0 (fixed) & $0.0009^{+0.0087}_{-0.0113}$  \\
$e\sin\omega$ & 0 (fixed) & $0.0007^{+0.0010}_{-0.0105}$ \\
$\chi^2_\mathrm{red}$ & 1.295 & \\
$\sigma_\mathrm{rms}$ (mmag) & 1.074 &
\enddata
\tablecomments{The \jktebop\ parameters quoted are median values and 68\% confidence intervals from 1000 Monte Carlo simulations. The MCMC parameters assumed a Gaussian prior on eccentricity with $\mu$=0.0, $\sigma$=0.01, as well as for the linear limb darkening parameter, with $\mu$=0.6, $\sigma$=0.1.}
\end{deluxetable}


\section{Test of Model Isochrones at Pleiades Age}
\label{sec:disc}

\begin{figure}
\centering
\includegraphics[width=0.75\textwidth]{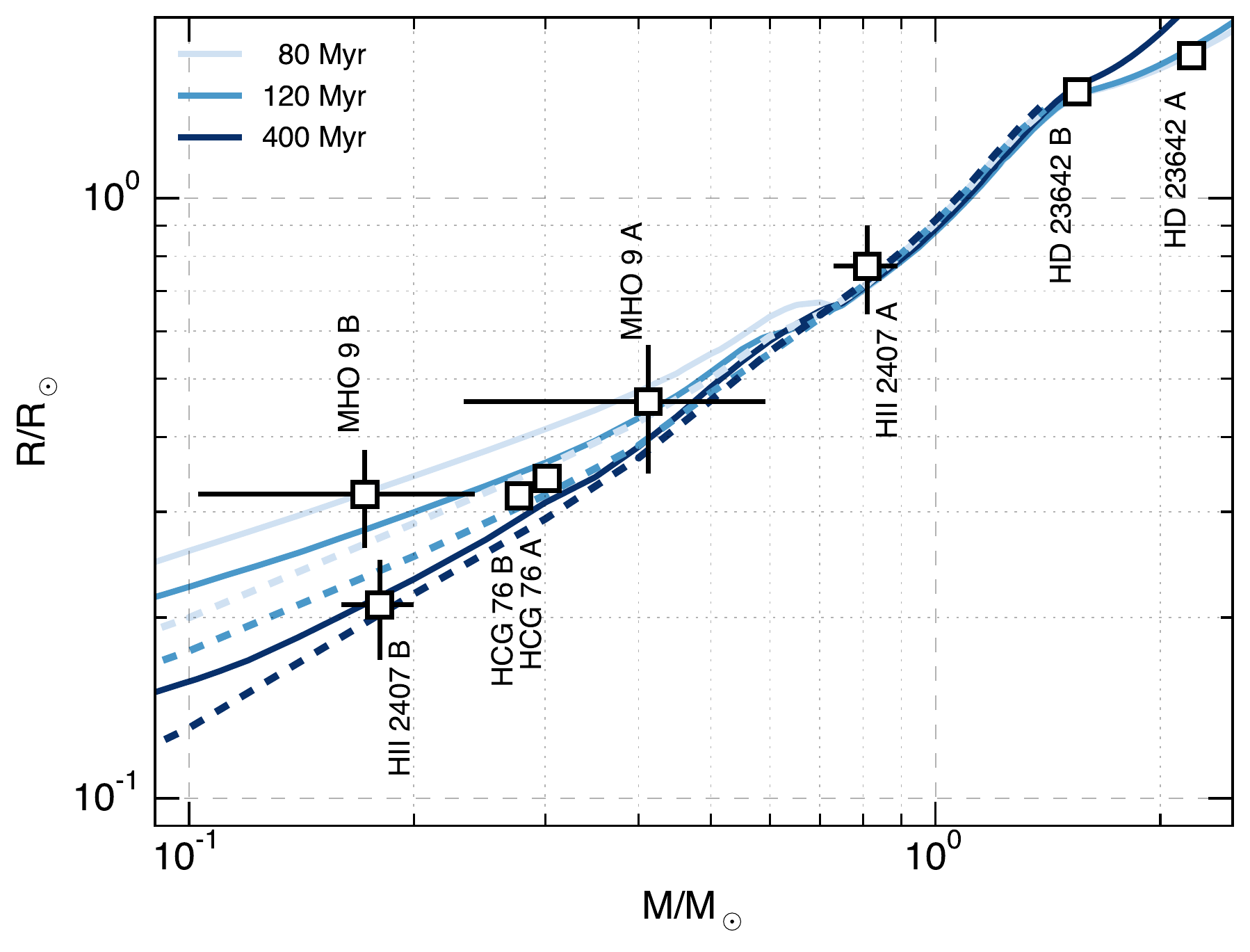}
\caption{Mass-radius diagram for all currently known eclipsing binaries in the Pleiades star cluster. With the exception of HII 2407 which is single-lined, each system is double-lined with dynamically determined masses. The solid and dashed curves show PARSEC v1.2S and BHAC15 isochrones for solar metallicity ($Z$=0.02).}
\label{fig:pleiades-mr}
\end{figure}

Our resulting models for each of the Pleiades EBs above allows us to examine all of the stellar components at once in a single mass-radius diagram, spanning a large range of masses, in comparison to the predictions of stellar evolution models.  Figure~\ref{fig:pleiades-mr} shows the mass-radius relations of all known Pleiades EB components, both previously published and reported here.  Neither the PARSEC v1.2S nor the BHAC15 isochrones extend across the entire mass range probed by these EBs, so we show both sets of isochrones at 80, 120, and 400 Myr for solar metallicity ($Z$=0.02).  We note again that the currently accepted age of the Pleiades is $125 \pm 8$ Myr \citep{stauffer1998} though with recent suggestions of a slightly younger age \citep[$112\pm5$ Myr;][]{dahm2015}.  

The three EB components at masses $\gtrsim0.5$ \msun\ appear to be largely consistent with both sets of isochrones at 120 Myr.  
As discussed in Sec.~\ref{sec:hd23642} and shown in Fig.~\ref{fig:hd23642-parsec}, our updated parameters for HD 23642 largely resolve discrepancies for this system from previous works. 

At low masses our measurements for the EB components agree better with the 120 Myr isochrone from BHAC15 than the same isochrone from PARSEC.  
This is most apparent for the lowest mass object, HII 2407B at a mass of $\sim$0.2 \msun, which clearly prefers the BHAC15 isochrone at 120 Myr and is inconsistent with the PARSEC isochrone at 120 Myr by 2--3$\sigma$. The exception at low masses is BPL 116B, however its mass uncertainty is large and therefore the discrepancy with the BHAC15 isochrone at 120 Myr is only $\sim$1$\sigma$. 

As noted in Sec.~\ref{sec:hcg76} and shown in Fig.~\ref{fig:hcg76bhac15}, our current best-fit constraints on HCG 76 suggest a modest preference for the slightly younger Pleiades age of \citet{dahm2015}. However, the radius ratio for HCG 76 is currently poorly constrained, and moreover the lowest mass component of HII 2407 is more consistent with the age of 120 Myr \citep{stauffer1998} than with the younger age of \citet{dahm2015}. Thus, overall the collective assessment of the Pleiades EBs spanning masses 0.2--2 \msun\ is to clearly prefer the BHAC15 models over the PARSEC models at a Pleiades age of $\approx$120 Myr.

\section{Summary and Conclusions}
\label{sec:summary}

We report the discovery of two new eclipsing binaries in the Pleiades cluster from inspection of K2 data, doubling the total number of known EBs in the cluster. These two systems have the lowest primary masses of the known EBs in the cluster, and thus all four of the stellar components are still in the pre-main-sequence phase of evolution at Pleiades age. With follow-up studies they may be elevated to benchmark status, becoming critically important anchors for evolution models that aim to reproduce both bulk stellar parameters and radiative properties at fixed ages. Importantly, both of the new EB systems have relatively large separations between components, reducing the likelihood that their individual properties are corrupted by interaction effects and increasing their value as calibrators.

As these two new Pleiades EBs have long orbital periods relative to the K2 campaign duration, the ephemerides of both still have large uncertainties. Thus, follow-up radial velocities and eclipse photometry are critically needed to better characterize these systems.
Even so, we have measured the masses of the components of HCG 76 to $\lesssim$2.5\% precision, the radii of these stars to $\lesssim$4.5\% precision, and have determined masses and radii for all four of the newly discovered Pleiades EB components. The highly precise parameters for HCG~76 further permit us to determine an independent Pleiades distance of 132$\pm$5 pc. In addition, we have newly measured the masses and radii of the components of the previously known Pleiades EB HD 23642, with updated stellar radii that largely resolve discrepant radii for this system relative to the Pleiades isochrone. Finally, together with a fourth Pleiades EB discovered in the K2 data and analyzed by \citet{david2015a}, we assess the overall agreement of stellar model isochrones at Pleiades ages over the mass range 0.2--2 \msun, finding broad agreement for the BHAC15 stellar models at an age of 120 Myr. 

In addition, we characterized a likely planetary mass companion in the Hyades, vA 50b, concurrently discovered and studied by \cite{mann2016}. We confirm the finding of those authors of a Neptune sized transiting object. With the additional radial velocity measurements presented here, we are able to improve the constraint on the maximum mass of the planet, yielding a maximum mass of 1.15\mjup. Extrasolar planets with well-constrained ages are extremely scarce, making this system valuable for constraining planet formation, migration, and evolution theories that aim to explain planetary and orbital parameters as a function of age. Most interesting about this planet is that it has a short orbital period and a relatively large size, which is particularly well constrained because of the host star's membership to a cluster with an extensively studied distance and age. Holding the radius fixed, we estimate that a planet mass of 1$M_\mathrm{Jup}$ would yield a bulk density an order of magnitude more dense than any of the terrestrial planets, while a mass of 1$M_\oplus$ would imply a density a factor of $\sim$4 less dense than Saturn. We consider either of these two extremes to be unlikely. In any case, the existence of such a large planet on a short orbital period at $\sim$600 Myr can be used in the future to place constraints on theories of planet formation and migration.

The upcoming K2 Campaign 13 will also target the Hyades cluster, allowing for the opportunity to discover new EBs and planets among members that were not included in the Campaign 4 pointing.


\acknowledgments
We thank the referee for helpful comments which led to significant improvement in the quality of this work. T.J.D.\ thanks J.\ Southworth for helpful discussions regarding the use of \jktebop, and I.\ Crossfield for providing his MCMC transit fitting routine which was used to fit vA 50b.
Support for this work was provided by NASA via grant NNX15AV62G. Some of the material presented herein is based upon work supported in 2015 by the National Science Foundation Graduate Research Fellowship under Grant DGE1144469. T.J.D.\ gratefully acknowledges research activities support from F.\ C\'{o}rdova through the 
Neugebauer Scholarship. Some of the data presented in this paper were obtained from the Mikulski Archive for Space Telescopes. STScI is operated by the Association of Universities for Research in Astronomy, Inc., under NASA contract NAS5-26555. Support for MAST for non-HST data is provided by the NASA Office of Space Science via grant NNX09AF08G and by other grants and contracts. This paper includes data collected by the Kepler mission. Funding for the Kepler mission is provided by the NASA Science Mission directorate. Some of the data presented herein were obtained at the W.M.\ Keck Observatory, which is operated as a scientific partnership among the California Institute of Technology, the University of California and the National Aeronautics and Space Administration. The Observatory was made possible by the generous financial support of the W.M.\ Keck Foundation. The authors wish to recognize and acknowledge the very significant cultural role and reverence that the summit of Mauna Kea has always had within the indigenous Hawaiian community.  We are most fortunate to have the opportunity to conduct observations from this mountain.

\bibliography{main}

\begin{thebibliography}{}
\expandafter\ifx\csname natexlab\endcsname\relax\def\natexlab#1{#1}\fi

\bibitem[{{Abt}(1958)}]{abt1958}
{Abt}, H.~A. 1958, \apj, 128, 139

\bibitem[{{Abt} \& {Levato}(1978)}]{abt1978}
{Abt}, H.~A., \& {Levato}, H. 1978, \pasp, 90, 201

\bibitem[{{Aigrain} {et~al.}(2015){Aigrain}, {Hodgkin}, {Irwin}, {Lewis}, \&
  {Roberts}}]{aigrain2015}
{Aigrain}, S., {Hodgkin}, S.~T., {Irwin}, M.~J., {Lewis}, J.~R., \& {Roberts},
  S.~J. 2015, \mnras, 447, 2880

\bibitem[{{Armstrong} {et~al.}(2015){Armstrong}, {Kirk}, {Lam}, {McCormac},
  {Walker}, {Brown}, {Osborn}, {Pollacco}, \& {Spake}}]{armstrong2015}
{Armstrong}, D.~J., {Kirk}, J., {Lam}, K.~W.~F., {et~al.} 2015, \aap, 579, A19

\bibitem[{{Artiukhina} \& {Kalinina}(1970)}]{ak1970}
{Artiukhina}, N.~M., \& {Kalinina}, E.~P. 1970, Trudy Gosudarstvennogo
  Astronomicheskogo Instituta, 39, 111

\bibitem[{{Baraffe} {et~al.}(2015){Baraffe}, {Homeier}, {Allard}, \&
  {Chabrier}}]{baraffe2015}
{Baraffe}, I., {Homeier}, D., {Allard}, F., \& {Chabrier}, G. 2015, \aap, 577,
  A42

\bibitem[{{Barrado y Navascues} \& {Stauffer}(1996)}]{barrado1996}
{Barrado y Navascues}, D., \& {Stauffer}, J.~R. 1996, \aap, 310, 879

\bibitem[{{Berta} {et~al.}(2013){Berta}, {Irwin}, \& {Charbonneau}}]{berta2013}
{Berta}, Z.~K., {Irwin}, J., \& {Charbonneau}, D. 2013, \apj, 775, 91

\bibitem[{{Bonfils} {et~al.}(2013){Bonfils}, {Delfosse}, {Udry}, {Forveille},
  {Mayor}, {Perrier}, {Bouchy}, {Gillon}, {Lovis}, {Pepe}, {Queloz}, {Santos},
  {S{\'e}gransan}, \& {Bertaux}}]{bonfils2013}
{Bonfils}, X., {Delfosse}, X., {Udry}, S., {et~al.} 2013, \aap, 549, A109

\bibitem[{{Bouy} {et~al.}(2015){Bouy}, {Bertin}, {Sarro}, {Barrado}, {Moraux},
  {Bouvier}, {Cuillandre}, {Berihuete}, {Olivares}, \& {Beletsky}}]{bouy2015}
{Bouy}, H., {Bertin}, E., {Sarro}, L.~M., {et~al.} 2015, \aap, 577, A148

\bibitem[{{Brandt} \& {Huang}(2015)}]{brandt2015}
{Brandt}, T.~D., \& {Huang}, C.~X. 2015, \apj, 807, 24

\bibitem[{{Bressan} {et~al.}(2012){Bressan}, {Marigo}, {Girardi}, {Salasnich},
  {Dal Cero}, {Rubele}, \& {Nanni}}]{bressan2012}
{Bressan}, A., {Marigo}, P., {Girardi}, L., {et~al.} 2012, \mnras, 427, 127

\bibitem[{{Brice{\~n}o} {et~al.}(1998){Brice{\~n}o}, {Hartmann}, {Stauffer}, \&
  {Mart{\'{\i}}n}}]{briceno1998}
{Brice{\~n}o}, C., {Hartmann}, L., {Stauffer}, J., \& {Mart{\'{\i}}n}, E. 1998,
  \aj, 115, 2074

\bibitem[{{Caffau} {et~al.}(2011){Caffau}, {Ludwig}, {Steffen}, {Freytag}, \&
  {Bonifacio}}]{caffau2011}
{Caffau}, E., {Ludwig}, H.-G., {Steffen}, M., {Freytag}, B., \& {Bonifacio}, P.
  2011, \solphys, 268, 255

\bibitem[{{Chen} {et~al.}(2015){Chen}, {Bressan}, {Girardi}, {Marigo}, {Kong},
  \& {Lanza}}]{chen2015}
{Chen}, Y., {Bressan}, A., {Girardi}, L., {et~al.} 2015, \mnras, 452, 1068

\bibitem[{{Claret} \& {Bloemen}(2011)}]{claret2011}
{Claret}, A., \& {Bloemen}, S. 2011, \aap, 529, A75

\bibitem[{{Claret} {et~al.}(2012){Claret}, {Hauschildt}, \&
  {Witte}}]{claret2012}
{Claret}, A., {Hauschildt}, P.~H., \& {Witte}, S. 2012, \aap, 546, A14

\bibitem[{{Conroy} {et~al.}(2014){Conroy}, {Pr{\v s}a}, {Stassun}, {Bloemen},
  {Parvizi}, {Quarles}, {Boyajian}, {Barclay}, {Shporer}, {Latham}, \&
  {Abdul-Masih}}]{conroy2014}
{Conroy}, K.~E., {Pr{\v s}a}, A., {Stassun}, K.~G., {et~al.} 2014, \pasp, 126,
  914

\bibitem[{{Crossfield} {et~al.}(2015){Crossfield}, {Petigura}, {Schlieder},
  {Howard}, {Fulton}, {Aller}, {Ciardi}, {L{\'e}pine}, {Barclay}, {de Pater},
  {de Kleer}, {Quintana}, {Christiansen}, {Schlafly}, {Kaltenegger}, {Crepp},
  {Henning}, {Obermeier}, {Deacon}, {Weiss}, {Isaacson}, {Hansen}, {Liu},
  {Greene}, {Howell}, {Barman}, \& {Mordasini}}]{crossfield2015}
{Crossfield}, I.~J.~M., {Petigura}, E., {Schlieder}, J.~E., {et~al.} 2015,
  \apj, 804, 10

\bibitem[{{Cutri} \& {et al.}(2012)}]{cutri2012}
{Cutri}, R.~M., \& {et al.} 2012, VizieR Online Data Catalog, 2311, 0

\bibitem[{{Dahm}(2015)}]{dahm2015}
{Dahm}, S.~E. 2015, \apj, 813, 108

\bibitem[{{David} \& {Hillenbrand}(2015)}]{david2015}
{David}, T.~J., \& {Hillenbrand}, L.~A. 2015, \apj, 804, 146

\bibitem[{{David} {et~al.}(2016){David}, {Hillenbrand}, {Cody}, {Carpenter}, \&
  {Howard}}]{david2016}
{David}, T.~J., {Hillenbrand}, L.~A., {Cody}, A.~M., {Carpenter}, J.~M., \&
  {Howard}, A.~W. 2016, \apj, 816, 21

\bibitem[{David {et~al.}(2015)David, Stauffer, Hillenbrand, Cody, Conroy,
  Stassun, Pope, Aigrain, Gillen, Cameron, Barrado, Rebull, Isaacson, Marcy,
  Zhang, Riddle, Ziegler, Law, \& Baranec}]{david2015a}
David, T.~J., Stauffer, J., Hillenbrand, L.~A., {et~al.} 2015, The
  Astrophysical Journal, 814, 62

\bibitem[{{de Bruijne} {et~al.}(2001){de Bruijne}, {Hoogerwerf}, \& {de
  Zeeuw}}]{debruijne2001}
{de Bruijne}, J.~H.~J., {Hoogerwerf}, R., \& {de Zeeuw}, P.~T. 2001, \aap, 367,
  111

\bibitem[{{Deacon} \& {Hambly}(2004)}]{deacon2004}
{Deacon}, N.~R., \& {Hambly}, N.~C. 2004, \aap, 416, 125

\bibitem[{{Delfosse} {et~al.}(2000){Delfosse}, {Forveille}, {S{\'e}gransan},
  {Beuzit}, {Udry}, {Perrier}, \& {Mayor}}]{delfosse2000}
{Delfosse}, X., {Forveille}, T., {S{\'e}gransan}, D., {et~al.} 2000, \aap, 364,
  217

\bibitem[{{Etzel}(1975)}]{etzel1975}
{Etzel}, P.~B. 1975, Master's thesis, Masters Thesis.~San Diego State
  University (1975)

\bibitem[{{Etzel}(1981)}]{etzel1981}
{Etzel}, P.~B. 1981, in Photometric and Spectroscopic Binary Systems, ed. E.~B.
  {Carling} \& Z.~{Kopal}, 111

\bibitem[{{Foreman-Mackey} {et~al.}(2013){Foreman-Mackey}, {Hogg}, {Lang}, \&
  {Goodman}}]{foreman-mackey2013}
{Foreman-Mackey}, D., {Hogg}, D.~W., {Lang}, D., \& {Goodman}, J. 2013, \pasp,
  125, 306

\bibitem[{{Goldman} {et~al.}(2013){Goldman}, {R{\"o}ser}, {Schilbach},
  {Magnier}, {Olczak}, {Henning}, {Juri{\'c}}, {Schlafly}, {Chen}, {Platais},
  {Burgett}, {Hodapp}, {Heasley}, {Kudritzki}, {Morgan}, {Price}, {Tonry}, \&
  {Wainscoat}}]{goldman2013}
{Goldman}, B., {R{\"o}ser}, S., {Schilbach}, E., {et~al.} 2013, \aap, 559, A43

\bibitem[{{Groenewegen} {et~al.}(2007){Groenewegen}, {Decin}, {Salaris}, \& {De
  Cat}}]{groenewegen2007}
{Groenewegen}, M.~A.~T., {Decin}, L., {Salaris}, M., \& {De Cat}, P. 2007,
  \aap, 463, 579

\bibitem[{{Guinan} \& {Ribas}(2001)}]{guinan2001}
{Guinan}, E.~F., \& {Ribas}, I. 2001, \apjl, 546, L43

\bibitem[{{Hambly} {et~al.}(1993){Hambly}, {Hawkins}, \&
  {Jameson}}]{hambly1993}
{Hambly}, N.~C., {Hawkins}, M.~R.~S., \& {Jameson}, R.~F. 1993, \aaps, 100, 607

\bibitem[{{Hanson}(1975)}]{hanson1975}
{Hanson}, R.~B. 1975, \aj, 80, 379

\bibitem[{{Haro} {et~al.}(1982){Haro}, {Chavira}, \& {Gonzalez}}]{haro1982}
{Haro}, G., {Chavira}, E., \& {Gonzalez}, G. 1982, Boletin del Instituto de
  Tonantzintla, 3, 3

\bibitem[{{Hauschildt} {et~al.}(1999){Hauschildt}, {Allard}, {Ferguson},
  {Baron}, \& {Alexander}}]{Hauschildt1999}
{Hauschildt}, P.~H., {Allard}, F., {Ferguson}, J., {Baron}, E., \& {Alexander},
  D.~R. 1999, \apj, 525, 871

\bibitem[{{H{\o}g} {et~al.}(2000){H{\o}g}, {Fabricius}, {Makarov}, {Urban},
  {Corbin}, {Wycoff}, {Bastian}, {Schwekendiek}, \& {Wicenec}}]{hog2000}
{H{\o}g}, E., {Fabricius}, C., {Makarov}, V.~V., {et~al.} 2000, \aap, 355, L27

\bibitem[{{Howell} {et~al.}(2014){Howell}, {Sobeck}, {Haas}, {Still},
  {Barclay}, {Mullally}, {Troeltzsch}, {Aigrain}, {Bryson}, {Caldwell},
  {Chaplin}, {Cochran}, {Huber}, {Marcy}, {Miglio}, {Najita}, {Smith},
  {Twicken}, \& {Fortney}}]{howell2014}
{Howell}, S.~B., {Sobeck}, C., {Haas}, M., {et~al.} 2014, \pasp, 126, 398

\bibitem[{{Jones} {et~al.}(1999){Jones}, {Fischer}, \& {Soderblom}}]{jones1999}
{Jones}, B.~F., {Fischer}, D., \& {Soderblom}, D.~R. 1999, \aj, 117, 330

\bibitem[{{Kamai} {et~al.}(2014){Kamai}, {Vrba}, {Stauffer}, \&
  {Stassun}}]{kamai2014}
{Kamai}, B.~L., {Vrba}, F.~J., {Stauffer}, J.~R., \& {Stassun}, K.~G. 2014,
  \aj, 148, 30

\bibitem[{{Kazarovets}(1993)}]{Kazarovets1993}
{Kazarovets}, E.~V. 1993, Peremennye Zvezdy, 23

\bibitem[{{Kennedy} {et~al.}(2007){Kennedy}, {Kenyon}, \&
  {Bromley}}]{kennedy2007}
{Kennedy}, G.~M., {Kenyon}, S.~J., \& {Bromley}, B.~C. 2007, \apss, 311, 9

\bibitem[{{Kirk} {et~al.}(2015){Kirk}, {Conroy}, {Pr{\v s}a}, {Abdul-Masih},
  {Kochoska}, {Matijevi{\v c}}, {Hambleton}, {Barclay}, {Bloemen}, {Boyajian},
  {Doyle}, {Fulton}, {Hoekstra}, {Jek}, {Kane}, {Kostov}, {Latham}, {Mazeh},
  {Orosz}, {Pepper}, {Quarles}, {Ragozzine}, {Shporer}, {Southworth},
  {Stassun}, {Thompson}, {Welsh}, {Agol}, {Derekas}, {Devor}, {Fischer},
  {Green}, {Gropp}, {Jacobs}, {Johnston}, {LaCourse}, {Saetre}, {Schwengeler},
  {Toczyski}, {Werner}, {Garrett}, {Gore}, {Martinez}, {Spitzer}, {Stevick},
  {Thomadis}, {Halley Vrijmoet}, {Yenawine}, {Batalha}, \&
  {Borucki}}]{kirk2016}
{Kirk}, B., {Conroy}, K., {Pr{\v s}a}, A., {et~al.} 2015, ArXiv e-prints,
  arXiv:1512.08830

\bibitem[{{LaCourse} {et~al.}(2015){LaCourse}, {Jek}, {Jacobs}, {Winarski},
  {Boyajian}, {Rappaport}, {Sanchis-Ojeda}, {Conroy}, {Nelson}, {Barclay},
  {Fischer}, {Schmitt}, {Wang}, {Stassun}, {Pepper}, {Coughlin}, {Shporer}, \&
  {Pr{\v s}a}}]{lacourse2015}
{LaCourse}, D.~M., {Jek}, K.~J., {Jacobs}, T.~L., {et~al.} 2015, \mnras, 452,
  3561

\bibitem[{{Laughlin} {et~al.}(2004){Laughlin}, {Bodenheimer}, \&
  {Adams}}]{laughlin2004}
{Laughlin}, G., {Bodenheimer}, P., \& {Adams}, F.~C. 2004, \apjl, 612, L73

\bibitem[{{Mann} {et~al.}(2015){Mann}, {Feiden}, {Gaidos}, {Boyajian}, \& {von
  Braun}}]{mann2015}
{Mann}, A.~W., {Feiden}, G.~A., {Gaidos}, E., {Boyajian}, T., \& {von Braun},
  K. 2015, \apj, 804, 64

\bibitem[{{Mann} {et~al.}(2016){Mann}, {Gaidos}, {Mace}, {Johnson}, {Bowler},
  {LaCourse}, {Jacobs}, {Vanderburg}, {Kraus}, {Kaplan}, \& {Jaffe}}]{mann2016}
{Mann}, A.~W., {Gaidos}, E., {Mace}, G.~N., {et~al.} 2016, ArXiv e-prints,
  arXiv:1512.00483

\bibitem[{{McClure}(1980)}]{mcclure1980}
{McClure}, R.~D. 1980, in Bulletin of the American Astronomical Society,
  Vol.~12, Bulletin of the American Astronomical Society, 867

\bibitem[{{McClure}(1982)}]{mcclure1982}
{McClure}, R.~D. 1982, \apj, 254, 606

\bibitem[{{Melis} {et~al.}(2014){Melis}, {Reid}, {Mioduszewski}, {Stauffer}, \&
  {Bower}}]{melis2014}
{Melis}, C., {Reid}, M.~J., {Mioduszewski}, A.~J., {Stauffer}, J.~R., \&
  {Bower}, G.~C. 2014, Science, 345, 1029

\bibitem[{{Mermilliod} {et~al.}(1997){Mermilliod}, {Bratschi}, \&
  {Mayor}}]{mermilliod1997}
{Mermilliod}, J.-C., {Bratschi}, P., \& {Mayor}, M. 1997, \aap, 320, 74

\bibitem[{{Mermilliod} {et~al.}(2009){Mermilliod}, {Mayor}, \&
  {Udry}}]{mermilliod2009}
{Mermilliod}, J.-C., {Mayor}, M., \& {Udry}, S. 2009, \aap, 498, 949

\bibitem[{{Mermilliod} {et~al.}(1992){Mermilliod}, {Rosvick}, {Duquennoy}, \&
  {Mayor}}]{mermilliod1992}
{Mermilliod}, J.-C., {Rosvick}, J.~M., {Duquennoy}, A., \& {Mayor}, M. 1992,
  \aap, 265, 513

\bibitem[{{Milone} \& {Schiller}(2013)}]{milone2013}
{Milone}, E.~F., \& {Schiller}, S.~J. 2013, in IAU Symposium, Vol. 289, IAU
  Symposium, ed. R.~{de Grijs}, 227--230

\bibitem[{{Munari} {et~al.}(2004){Munari}, {Dallaporta}, {Siviero}, {Soubiran},
  {Fiorucci}, \& {Girard}}]{munari2004}
{Munari}, U., {Dallaporta}, S., {Siviero}, A., {et~al.} 2004, \aap, 418, L31

\bibitem[{{Nelson} \& {Davis}(1972)}]{nd1972}
{Nelson}, B., \& {Davis}, W.~D. 1972, \apj, 174, 617

\bibitem[{{Nidever} {et~al.}(2002){Nidever}, {Marcy}, {Butler}, {Fischer}, \&
  {Vogt}}]{nidever2002}
{Nidever}, D.~L., {Marcy}, G.~W., {Butler}, R.~P., {Fischer}, D.~A., \& {Vogt},
  S.~S. 2002, \apjs, 141, 503

\bibitem[{{{\"O}berg} {et~al.}(2011){{\"O}berg}, {Murray-Clay}, \&
  {Bergin}}]{oberg2011}
{{\"O}berg}, K.~I., {Murray-Clay}, R., \& {Bergin}, E.~A. 2011, \apjl, 743, L16

\bibitem[{{Pearce}(1957)}]{pearce1957}
{Pearce}, J.~A. 1957, Publications of the Dominion Astrophysical Observatory
  Victoria, 10, 435

\bibitem[{{Pecaut} \& {Mamajek}(2013)}]{pecaut2013}
{Pecaut}, M.~J., \& {Mamajek}, E.~E. 2013, \apjs, 208, 9

\bibitem[{{Perryman} {et~al.}(1998){Perryman}, {Brown}, {Lebreton}, {Gomez},
  {Turon}, {Cayrel de Strobel}, {Mermilliod}, {Robichon}, {Kovalevsky}, \&
  {Crifo}}]{perryman1998}
{Perryman}, M.~A.~C., {Brown}, A.~G.~A., {Lebreton}, Y., {et~al.} 1998, \aap,
  331, 81

\bibitem[{{Pinfield} {et~al.}(2000){Pinfield}, {Hodgkin}, {Jameson},
  {Cossburn}, {Hambly}, \& {Devereux}}]{pinfield2000}
{Pinfield}, D.~J., {Hodgkin}, S.~T., {Jameson}, R.~F., {et~al.} 2000, \mnras,
  313, 347

\bibitem[{{Popper} \& {Etzel}(1981)}]{popper1981}
{Popper}, D.~M., \& {Etzel}, P.~B. 1981, \aj, 86, 102

\bibitem[{{Pr{\v s}a} {et~al.}(2011){Pr{\v s}a}, {Batalha}, {Slawson}, {Doyle},
  {Welsh}, {Orosz}, {Seager}, {Rucker}, {Mjaseth}, {Engle}, {Conroy},
  {Jenkins}, {Caldwell}, {Koch}, \& {Borucki}}]{prsa2011}
{Pr{\v s}a}, A., {Batalha}, N., {Slawson}, R.~W., {et~al.} 2011, \aj, 141, 83

\bibitem[{{Reid}(1993)}]{reid1993}
{Reid}, N. 1993, \mnras, 265, 785

\bibitem[{{Roeser} {et~al.}(2010){Roeser}, {Demleitner}, \&
  {Schilbach}}]{roeser2010}
{Roeser}, S., {Demleitner}, M., \& {Schilbach}, E. 2010, \aj, 139, 2440

\bibitem[{{R{\"o}ser} {et~al.}(2011){R{\"o}ser}, {Schilbach}, {Piskunov},
  {Kharchenko}, \& {Scholz}}]{roser2011}
{R{\"o}ser}, S., {Schilbach}, E., {Piskunov}, A.~E., {Kharchenko}, N.~V., \&
  {Scholz}, R.-D. 2011, \aap, 531, A92

\bibitem[{{Sarro} {et~al.}(2014){Sarro}, {Bouy}, {Berihuete}, {Bertin},
  {Moraux}, {Bouvier}, {Cuillandre}, {Barrado}, \& {Solano}}]{sarro2014}
{Sarro}, L.~M., {Bouy}, H., {Berihuete}, A., {et~al.} 2014, \aap, 563, A45

\bibitem[{{Schiller} \& {Milone}(1987)}]{schiller1987}
{Schiller}, S.~J., \& {Milone}, E.~F. 1987, \aj, 93, 1471

\bibitem[{{Slawson} {et~al.}(2011){Slawson}, {Pr{\v s}a}, {Welsh}, {Orosz},
  {Rucker}, {Batalha}, {Doyle}, {Engle}, {Conroy}, {Coughlin}, {Gregg},
  {Fetherolf}, {Short}, {Windmiller}, {Fabrycky}, {Howell}, {Jenkins}, {Uddin},
  {Mullally}, {Seader}, {Thompson}, {Sanderfer}, {Borucki}, \&
  {Koch}}]{slawson2011}
{Slawson}, R.~W., {Pr{\v s}a}, A., {Welsh}, W.~F., {et~al.} 2011, \aj, 142, 160

\bibitem[{{Soderblom} {et~al.}(1999){Soderblom}, {King}, {Siess}, {Jones}, \&
  {Fischer}}]{soderblom1999}
{Soderblom}, D.~R., {King}, J.~R., {Siess}, L., {Jones}, B.~F., \& {Fischer},
  D. 1999, \aj, 118, 1301

\bibitem[{{Soderblom} {et~al.}(2009){Soderblom}, {Laskar}, {Valenti},
  {Stauffer}, \& {Rebull}}]{soderblom2009}
{Soderblom}, D.~R., {Laskar}, T., {Valenti}, J.~A., {Stauffer}, J.~R., \&
  {Rebull}, L.~M. 2009, \aj, 138, 1292

\bibitem[{{Southworth}(2012)}]{southworth2012}
{Southworth}, J. 2012, \mnras, 426, 1291

\bibitem[{{Southworth}(2013)}]{southworth2013}
---. 2013, \aap, 557, A119

\bibitem[{{Southworth} {et~al.}(2005){Southworth}, {Maxted}, \&
  {Smalley}}]{southworth2005}
{Southworth}, J., {Maxted}, P.~F.~L., \& {Smalley}, B. 2005, \aap, 429, 645

\bibitem[{{Stassun} {et~al.}(2014){Stassun}, {Feiden}, \&
  {Torres}}]{stassun2014}
{Stassun}, K.~G., {Feiden}, G.~A., \& {Torres}, G. 2014, \nar, 60, 1

\bibitem[{{Stauffer} {et~al.}(1991){Stauffer}, {Klemola}, {Prosser}, \&
  {Probst}}]{stauffer1991}
{Stauffer}, J., {Klemola}, A., {Prosser}, C., \& {Probst}, R. 1991, \aj, 101,
  980

\bibitem[{{Stauffer} {et~al.}(1998){Stauffer}, {Schultz}, \&
  {Kirkpatrick}}]{stauffer1998}
{Stauffer}, J.~R., {Schultz}, G., \& {Kirkpatrick}, J.~D. 1998, \apjl, 499,
  L199

\bibitem[{{Stauffer} {et~al.}(2007){Stauffer}, {Hartmann}, {Fazio}, {Allen},
  {Patten}, {Lowrance}, {Hurt}, {Rebull}, {Cutri}, {Ramirez}, {Young}, {Rieke},
  {Gorlova}, {Muzerolle}, {Slesnick}, \& {Skrutskie}}]{stauffer2007}
{Stauffer}, J.~R., {Hartmann}, L.~W., {Fazio}, G.~G., {et~al.} 2007, \apjs,
  172, 663

\bibitem[{{Torres}(2003)}]{torres2003}
{Torres}, G. 2003, Information Bulletin on Variable Stars, 5402, 1

\bibitem[{{Upgren} \& {Weis}(1977)}]{upgren1977}
{Upgren}, A.~R., \& {Weis}, E.~W. 1977, \aj, 82, 978

\bibitem[{{Upgren} {et~al.}(1985{\natexlab{a}}){Upgren}, {Weis}, \&
  {Hanson}}]{weis1985}
{Upgren}, A.~R., {Weis}, E.~W., \& {Hanson}, R.~B. 1985{\natexlab{a}}, \aj, 90,
  2039

\bibitem[{{Upgren} {et~al.}(1985{\natexlab{b}}){Upgren}, {Weis}, \&
  {Hanson}}]{upgren1985}
---. 1985{\natexlab{b}}, \aj, 90, 2039

\bibitem[{{Vaccaro} {et~al.}(2015){Vaccaro}, {Wilson}, {Van Hamme}, \&
  {Terrell}}]{vaccaro2015}
{Vaccaro}, T.~R., {Wilson}, R.~E., {Van Hamme}, W., \& {Terrell}, D. 2015,
  \apj, 810, 157

\bibitem[{{van Altena}(1966)}]{vanaltena1966}
{van Altena}, W.~F. 1966, \aj, 71, 482

\bibitem[{{van Altena}(1969)}]{vanaltena1969}
---. 1969, \aj, 74, 2

\bibitem[{{van Bueren}(1952)}]{vanbueren1952}
{van Bueren}, H.~G. 1952, \bain, 11, 385

\bibitem[{{van Leeuwen}(2009)}]{vanleeuwen2009}
{van Leeuwen}, F. 2009, \aap, 497, 209

\bibitem[{{Vanderburg} \& {Johnson}(2014)}]{vanderburg2014}
{Vanderburg}, A., \& {Johnson}, J.~A. 2014, \pasp, 126, 948

\bibitem[{{Vogt} {et~al.}(1994){Vogt}, {Allen}, {Bigelow}, {Bresee}, {Brown},
  {Cantrall}, {Conrad}, {Couture}, {Delaney}, {Epps}, {Hilyard}, {Hilyard},
  {Horn}, {Jern}, {Kanto}, {Keane}, {Kibrick}, {Lewis}, {Osborne},
  {Pardeilhan}, {Pfister}, {Ricketts}, {Robinson}, {Stover}, {Tucker}, {Ward},
  \& {Wei}}]{vogt1994}
{Vogt}, S.~S., {Allen}, S.~L., {Bigelow}, B.~C., {et~al.} 1994, in Society of
  Photo-Optical Instrumentation Engineers (SPIE) Conference Series, Vol. 2198,
  Instrumentation in Astronomy VIII, ed. D.~L. {Crawford} \& E.~R. {Craine},
  362

\bibitem[{{Weis}(1983)}]{weis1983}
{Weis}, E.~W. 1983, \pasp, 95, 29

\bibitem[{{Weis} {et~al.}(1979){Weis}, {Deluca}, \& {Upgren}}]{weis1979}
{Weis}, E.~W., {Deluca}, E.~E., \& {Upgren}, A.~R. 1979, \pasp, 91, 766

\bibitem[{{Weis} \& {Hanson}(1988)}]{weis1988}
{Weis}, E.~W., \& {Hanson}, R.~B. 1988, \aj, 96, 148

\bibitem[{{Zacharias} {et~al.}(2013){Zacharias}, {Finch}, {Girard}, {Henden},
  {Bartlett}, {Monet}, \& {Zacharias}}]{zacharias2013}
{Zacharias}, N., {Finch}, C.~T., {Girard}, T.~M., {et~al.} 2013, \aj, 145, 44

\bibitem[{{Zacharias} {et~al.}(2015){Zacharias}, {Finch}, {Subasavage},
  {Bredthauer}, {Crockett}, {Divittorio}, {Ferguson}, {Harris}, {Harris},
  {Henden}, {Kilian}, {Munn}, {Rafferty}, {Rhodes}, {Schultheiss}, {Tilleman},
  \& {Wieder}}]{zacharias2015}
{Zacharias}, N., {Finch}, C., {Subasavage}, J., {et~al.} 2015, \aj, 150, 101

\end{thebibliography}

\appendix 
\section{Model results for Pleiades non-member EB AK II 465}
\label{sec:akiiappendix} 

\begin{figure}[ht!]
\centering
\includegraphics[width=0.7\textwidth]{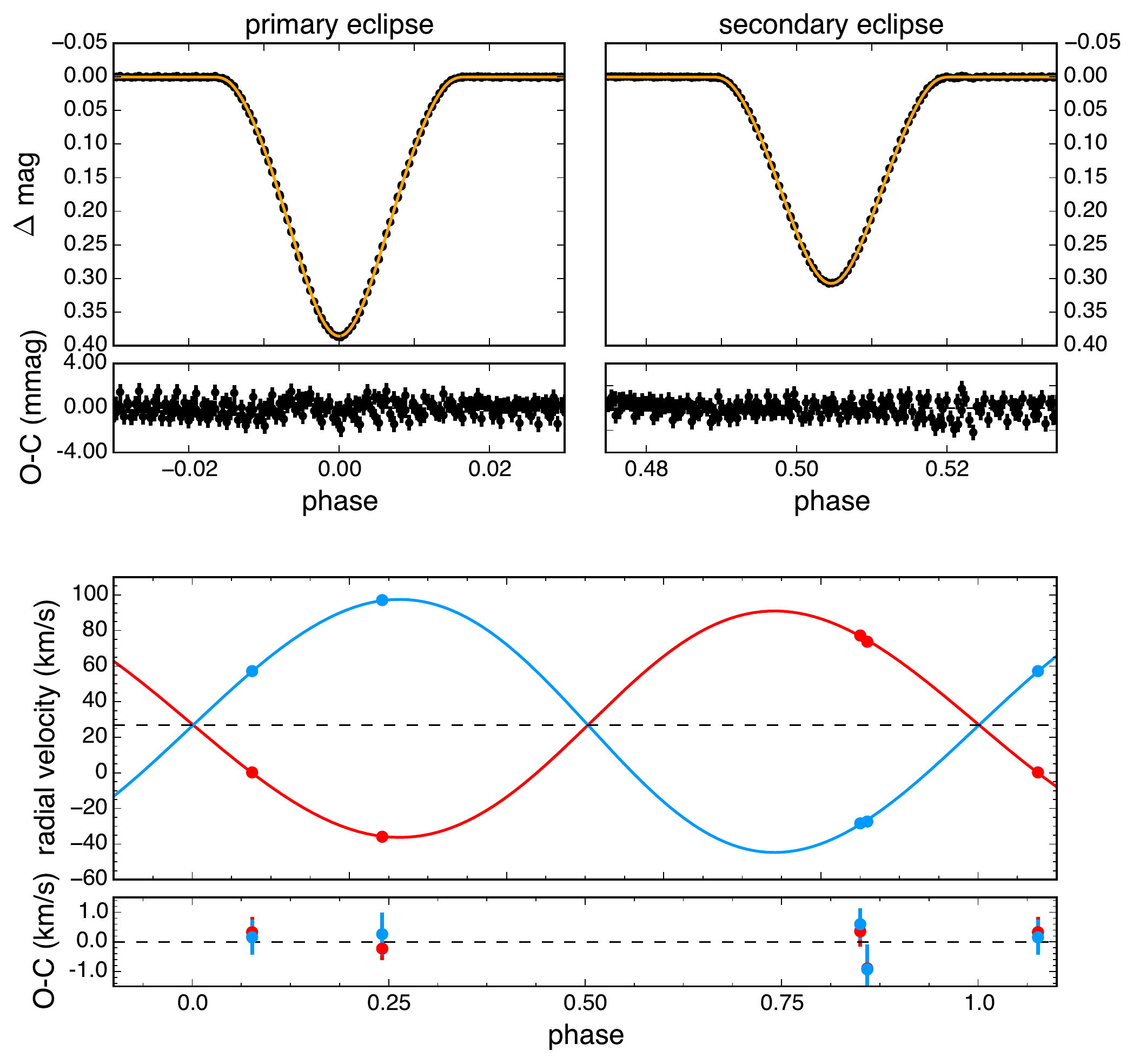}
\caption{Top panels: $K2$ PDC SAP light curve for AK II 465 phase folded on the orbital period, with the best-fit \jktebop\ model plotted in orange. Bottom panel: Radial velocities with the best-fit \jktebop\ models indicated by the red and blue curves. In each panel the best-fit residuals are plotted below. The structure in the residuals to the fit of the primary eclipse is likely due to inadequate modeling of limb darkening.}
\label{fig:akii465model}
\end{figure}

The best-fit parameters of the EB AK II 465 are summarized in Table~\ref{tab:akii465params} and the solution displayed in Figure~\ref{fig:akii465model}. 

\renewcommand{\period}{8.0746423}
\renewcommand{\eperiod}{0.0000067}
\renewcommand{\tzero}{72.234969}
\renewcommand{\etzero}{0.000030}
\renewcommand{\sbratio}{0.760}
\renewcommand{\esbratio}{0.014}
\renewcommand{\sumradii}{0.10077}
\renewcommand{\esumradii}{0.00038}
\renewcommand{\ratradii}{0.797}
\renewcommand{\eratradii}{0.016}
\renewcommand{\ldprim}{0.541}
\renewcommand{\eldprim}{0.037}
\renewcommand{\ldsec}{0.438}
\renewcommand{\eldsec}{0.052}
\renewcommand{\inclination}{87.987}
\renewcommand{\einclination}{0.034}
\renewcommand{\ecosw}{0.0072303}
\renewcommand{\eecosw}{0.0000069}
\renewcommand{\esinw}{-0.0358}
\renewcommand{\eesinw}{0.0018}
\renewcommand{\rvprim}{63.61}
\renewcommand{\ervprim}{0.31}
\renewcommand{\rvsec}{71.09}
\renewcommand{\ervsec}{0.44}
\renewcommand{\rvsys}{26.89}
\renewcommand{\ervsys}{0.19}
\renewcommand{\rada}{0.05608}
\renewcommand{\erada}{0.00035}
\renewcommand{\radb}{0.04469}
\renewcommand{\eradb}{0.00064}
\renewcommand{\lightratio}{0.503}
\renewcommand{\elightratio}{0.019}
\renewcommand{\ecc}{0.0365}
\renewcommand{\eecc}{0.0017}
\renewcommand{\omegap}{281.42}
\renewcommand{\eomegap}{0.55}
\renewcommand{\bpri}{0.649}
\renewcommand{\ebpri}{0.014}
\renewcommand{\bsec}{0.604}
\renewcommand{\ebsec}{0.014}
\renewcommand{\rchisq}{1.078}
\renewcommand{\rmslc}{0.720}
\renewcommand{\rmsprv}{0.521}
\renewcommand{\rmssrv}{0.572}
\renewcommand{\sma}{21.488} 
\renewcommand{\esma}{0.088} 
\renewcommand{\mrat}{0.8947} 
\renewcommand{\emrat}{0.0067} 
\renewcommand{\mpri}{1.079} 
\renewcommand{\empri}{0.015} 
\renewcommand{\msec}{0.965} 
\renewcommand{\emsec}{0.011} 
\renewcommand{\rpri}{1.2051} 
\renewcommand{\erpri}{0.0097} 
\renewcommand{\rsec}{0.960} 
\renewcommand{\ersec}{0.014} 
\renewcommand{\logga}{4.3087} 
\renewcommand{\elogga}{0.0058} 
\renewcommand{\loggb}{4.458} 
\renewcommand{\eloggb}{0.013} 
\renewcommand{\rhoa}{0.617} 
\renewcommand{\erhoa}{0.011} 
\renewcommand{\rhob}{1.090} 
\renewcommand{\erhob}{0.047} 

\begin{deluxetable}{lrll}
\tabletypesize{\scriptsize}
\tablecaption{ System Parameters of AK II 465 \label{tab:akii465params}}
\tablewidth{0.99\textwidth}
\tablehead{
\colhead{Parameter} & \colhead{Symbol} & \colhead{\jktebop} & \colhead{Units}
\\
& & \colhead{Value} &
} 

\startdata
Orbital period & $P$ & \period\ $\pm$ \eperiod & days \\
Ephemeris timebase - 2457000 & $T_0$ & \tzero\ $\pm$ \etzero & BJD \\
Surface brightness ratio & $J$ & \sbratio\ $\pm$ \esbratio &  \\
Sum of fractional radii & $(R_1+R_2)/a$ & \sumradii\ $\pm$ \esumradii & \\
Ratio of radii & $k$ & \ratradii\ $\pm$ \eratradii  & \\
Orbital inclination & $i$ & \inclination\ $\pm$ \einclination & deg \\
Primary limb darkening coefficient & $u_1$ & \ldprim $\pm$ \eldprim &  \\
Secondary limb darkening coefficient & $u_2$ & \ldsec $\pm$ \eldsec &  \\
Combined eccentricity, periastron longitude & $e\cos\omega$ & \ecosw\ $\pm$ \eecosw & \\
Combined eccentricity, periastron longitude & $e\sin\omega$ & \esinw\ $\pm$ \eesinw & \\
Primary radial velocity amplitude & $K_1$ & \rvprim\ $\pm$ \ervprim & km s$^{-1}$ \\
Secondary radial velocity amplitude & $K_2$ & \rvsec\ $\pm$ \ervsec & km s$^{-1}$ \\
Systemic radial velocity & $\gamma$ & \rvsys\ $\pm$ \ervsys & km s$^{-1}$ \\
Fractional radius of primary & $R_1/a$ & \rada $\pm$ \erada & \\
Fractional radius of secondary & $R_2/a$ & \radb $\pm$ \eradb & \\
Luminosity ratio & $L_2/L_1$ & \lightratio\ $\pm$ \elightratio & \\
Eccentricity & $e$ & \ecc\ $\pm$ \eecc & \\
Periastron longitude & $\omega$ & \omegap\ $\pm$ \eomegap & deg \\
Impact parameter of primary eclipse & $b_1$ & \bpri\ $\pm$ \ebpri & \\
Impact parameter of secondary eclipse & $b_2$ & \bsec\ $\pm$ \ebsec & \\
Orbital semi-major axis & $a$ & \sma\ $\pm$ \esma & \rsun  \\
Mass ratio & $q$ & \mrat\ $\pm$ \emrat & \\
Primary mass & $M_1$ & \mpri\ $\pm$ \empri & \msun \\
Secondary mass & $M_2$ & \msec\ $\pm$ \emsec & \msun \\
Primary radius & $R_1$ & \rpri\ $\pm$ \erpri & \rsun \\
Secondary radius & $R_2$ & \rsec\ $\pm$ \ersec & \rsun \\
Primary surface gravity & $\log g_1$ & \logga\ $\pm$ \elogga & cgs \\
Secondary surface gravity & $\log g_2$ & \loggb\ $\pm$ \eloggb & cgs \\
Primary mean density & $\rho_1$ & \rhoa\ $\pm$ \erhoa & $\rho_\odot$ \\
Secondary mean density & $\rho_2$ & \rhob\ $\pm$ \erhob & $\rho_\odot$ \\
Reduced chi-squared of light curve fit & $\chi^2_\mathrm{red}$ & \rchisq  & \\
RMS of best fit light curve residuals & & \rmslc &  mmag \\
Reduced chi-squared of primary RV fit & $\chi^2_\mathrm{red}$ & 0.63  &  \\
RMS of primary RV residuals & & \rmsprv & km s$^{-1}$ \\
Reduced chi-squared of secondary RV fit & $\chi^2_\mathrm{red}$ & 0.95  & \\
RMS of secondary RV residuals & & \rmssrv & km s$^{-1}$
\tablecomments{Best-fit orbital parameters and their uncertainties resulting from 1,000 Monte Carlo simulations with \jktebop.}
\enddata

\end{deluxetable}

\end{document}